\documentclass[%
 reprint,
 amsmath,amssymb,
 aps,
]{revtex4-2}

\usepackage[citecolor=blue, urlcolor=blue, linkcolor=blue, colorlinks=true]{hyperref}
\usepackage{graphicx}%
\usepackage{dcolumn}%
\usepackage{bm}%

\usepackage{mathtools}
\usepackage{braket}
\usepackage{graphicx}
\usepackage{color} %
\usepackage[dvipsnames]{xcolor}

\definecolor{Mycolor1}{cmyk}{0.1, 0.7808, 0.4429, 0.1412}
\definecolor{mygreen}{rgb}{0.35, 0.5, 0.0}

\usepackage{dsfont}
\usepackage{soul}
\usepackage{enumitem}

\begin{document}

\preprint{APS/123-QED}

\title{Scaling of neural-network quantum states for time evolution}

\author{Sheng-Hsuan Lin}
\affiliation{Department of Physics, TFK, Technische Universit{\"a}t M{\"u}nchen, James-Franck-Stra{\ss}e 1, 85748 Garching, Germany}
\author{Frank Pollmann}
\affiliation{Department of Physics, TFK, Technische Universit{\"a}t M{\"u}nchen, James-Franck-Stra{\ss}e 1, 85748 Garching, Germany}

\date{\today}%

\begin{abstract}
Simulating quantum many-body dynamics on classical computers is a challenging problem due to the exponential growth of the Hilbert space.
Artificial neural networks have recently been introduced as a new tool to approximate quantum-many body states.
We benchmark the variational power of the restricted Boltzmann machine quantum states and different shallow and deep neural autoregressive quantum states to simulate global quench dynamics of a non-integrable quantum Ising chain.
We find that the number of parameters required to represent the quantum state at a given accuracy increases exponentially in time.
The growth rate is only slightly affected by the network architecture over a wide range of different design choices: shallow and deep networks, small and large filter sizes, dilated and normal convolutions, with and without shortcut connections.
\end{abstract}

\maketitle

\section{\label{sec:intro} Introduction }
Numerical simulations are pivotal for the understanding of the emergent physics of quantum-many body systems.
Of particular current interest are quantum many-body systems out of equilibrium, where several exciting phenomena have recently been discovered such as many-body localization~\cite{basko2006metal, pal2010many, schreiber2015observation,choi2016exploring,smith2016many} and quantum many-body scars~\cite{bernien2017probing,turner2018weak,choi2019emergent}.
The interest in such systems is driven by rapid experimental progress, e.g., cold atoms and trapped ions allow for an unprecedented level of control and long coherence times ~\cite{greiner2002collapse,RevModPhys.80.885,RevModPhys.83.863,garttner2017measuring}.

Over the past decades, a large arsenal of numerical tools has been introduced to tackle such systems.
Widely used are exact diagonalization (ED) and Krylov-subspace methods, which simulate time evolution exactly.
The system size, however, is strongly limited by the exponential growth of the many-body Hilbert space.
In many cases, the size limitations can be remedied by using tensor network (TN) methods~\cite{verstraete2008matrix,schollwock2011density,bridgeman2017hand,hauschild2018efficient,hackbusch2012tensor}.
These include matrix-product states (MPS)~\cite{fannes1992abundance,perez2006matrix,perez2007matrix} methods for one-dimensional (1D) systems, which represent powerful numerical methods that are not limited by the system size but instead by the entanglement.
While ground states of local gapped Hamiltonians~\cite{white1992density} satisfy the entanglement area-law~\cite{eisert2010colloquium,hastings2007area} and can thus be efficiently simulated, the application of MPS to study non-equilibrium dynamics~\cite{white2004real,daley2004time,vidal2003efficient,vidal2004efficient,10.21468/SciPostPhys.8.2.021,PhysRevB.98.045110} is strongly limited by the fast growth of entanglement.
In particular, the entanglement generically grows linearly and hence leads to an exponential growth of the required parameters.

A promising new approach to simulate the dynamics of many-body systems is based on a representation of quantum states using artificial neural networks (ANNs)~\cite{carleo2017solving}.
Recent works have shown not only promising results for simulating non-equilibrium dynamics~\cite{schmitt2020quantum}, but also in describing two-dimensional (2D) critical ground states~\cite{sharir2020deep} and states with chiral topological order~\cite{PhysRevX.8.011006}.
Importantly, states based on ANNs can efficiently encode volume law entanglement~\cite{deng2017quantum,levine2019quantum} and are thus per se not limited by the entanglement growth in non-equilibrium systems.
It remains an open question what limits the representation based on ANNs and how the number of parameters generically scales.

\begin{figure}[b]
\centering
\includegraphics[width =0.7\columnwidth]{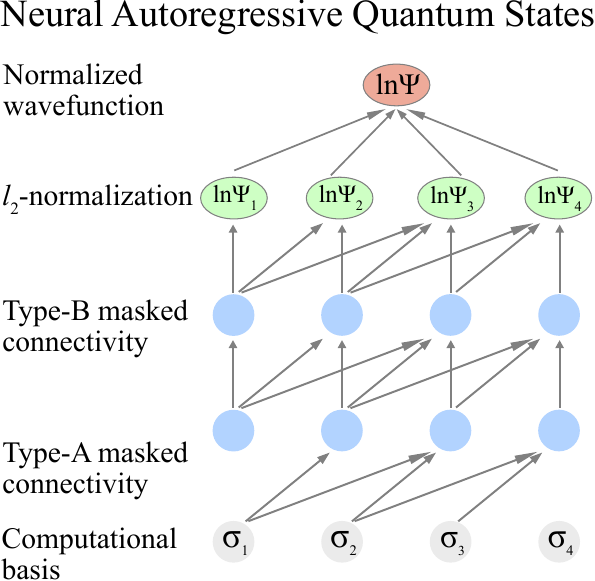}
\caption{ Illustration of a neural autoregressive quantum state. The neural network approximates the conditional probability amplitudes $\Psi_1, \Psi_2,\ldots $ in the log space.
Each of the conditional probability amplitudes are normalized to give a normalized wavefunction.
The network connectivity is restricted (masked) for causality, such that $\Psi_j$ only depends on $\sigma_{1}\ldots \sigma_{j-1}$.
The network itself could be either a fully-connected neural network or a convolutional neural network.
\label{fig: ansatz}}
\end{figure}

In this paper, we benchmark the variational power of a number of different ANNs approaches to simulate quench dynamics in 1D and compare to MPS based approximations. 
We find numerical evidence that states based on ANNs can describe quantum many-body states under quenched time evolution of local Hamiltonian with a number of parameters that grows exponentially in time---which in fact resembles the scaling of MPS based simulations.
The remainder of the paper is as follows:
We first describe the variational wavefunctions based on ANNs and the variational method to approximate quantum states using supervised learning in Sec.~\ref{sec: methods}.
To study the time scales accessible by different variational wavefunctions, we variationally optimize wavefunctions to approximate the exact time evolution in Section~\ref{sec: te}.
In Section~\ref{sec: conclusion}, We conclude by discussing the implication on numerical algorithms and possible future directions.

\section{ Methods \label{sec: methods}}

In this section, we review variational wavefunctions based on general neural networks and autoregressive neural networks.
Moreover, we introduce a novel cost function to approximate complex-valued quantum states with autoregressive neural networks in the supervised learning framework~\cite{lecun2015deep,goodfellow2016deep}.

\subsection{ Neural-Network Quantum States \label{subsec: NQS}}

A pure quantum state $\ket{\Psi}$ describing a quantum many-body system in the Hilbert space $\mathcal{H}=\otimes_{i}^N \mathcal{H}_i$ can be represented as
\begin{align*}
    \ket{\Psi}&= \sum_{\sigma_1,\sigma_2 \cdots ,\sigma_N} \Psi_{\sigma_1 \sigma_2 \cdots \sigma_N} \ket{ \sigma_1} \otimes \ket{ \sigma_2} \otimes \cdots \otimes \ket{\sigma_N},
\end{align*}
where $\{\mathcal{H}_i\}$ and $\ket{\sigma_i}$ are the local Hilbert spaces and the corresponding local basis states, respectively. 
The exact representation $\Psi_{\sigma_1 \sigma_2 \cdots \sigma_N}$ requires a number of parameters that grows exponentially with the system size $N$.
Neural-network quantum states (NQS)~\cite{carleo2017solving} are variational wavefunctions that approximate the amplitudes by neural networks $f$ with a polynomial number of parameters $\text{poly}(N)$,
\begin{equation}
\Psi_{\sigma_1 \sigma_2 \cdots \sigma_N} \approx f(\sigma_1, \sigma_2, \cdots ,\sigma_N ; \bm{w} )     \coloneqq \Psi^\text{NN}_{\sigma_1 \sigma_2 \cdots \sigma_N},
\end{equation}
where $\bm{w}$ are the parameters. 
One of the motivations to consider neural networks is the universal approximation theorem~\cite{hornik1989multilayer,cybenko1989approximation,leshno1993multilayer,kim2002universal,voigtlaender2020universal}, which guarantees an arbitrarily close approximation provided a sufficiently large number of hidden neurons is included (a review of neural networks can be found in Appendix~\ref{appendix: NN}).
Various different types of neural networks have been considered for approximating quantum states, including restricted Boltzmann machines (RBMs)~\cite{{hinton2006reducing,salakhutdinov2007restricted}} in Ref.~\onlinecite{carleo2017solving}, fully-connected feed-forward neural networks (NNs) in Ref.~\onlinecite{cai2018approximating}, convolutional neural networks (CNNs) in Ref.~\onlinecite{PhysRevB.100.125124}, and recurrent neural networks (RNNs)~\cite{rumelhart1986learning,hochreiter1997long} in Refs.~\onlinecite{hibat2020recurrent,luo2020probabilistic}.
NQS found applications in studying ground states and low-energy states~\cite{choo2018symmetries,choo2020fermionic,nomura2021helping,PhysRevLett.124.097201,vieijra2021many}, quantum state tomography~\cite{carrasquilla2019reconstructing}, simulating open quantum systems~\cite{PhysRevLett.122.250502,PhysRevLett.122.250503,luo2020probabilistic,reh2021time,PhysRevLett.122.250501,PhysRevB.99.214306}, and real-time evolution~\cite{carleo2017solving,schmitt2020quantum,gutierrez2019real,10.21468/SciPostPhys.7.1.004,fabiani2021supermagnonic,PhysRevB.98.024311}.

In general, NQS have the following properties:
The unnormalized probability amplitude and its gradient $\nabla_{\bm{w}} \Psi^{NN}$ can be efficiently evaluated at a cost of $\mathcal{O}(N^2)$.
This property allows one to combine NQS with conventional Monte Carlo methods.
For example, the expectation value of  few-body observables $\hat{O}$ can be computed as 
\begin{equation}
    \frac{\braket{ \Psi^{NN} | \hat{O} | \Psi^{NN} }}{\braket{ \Psi^{NN} |  \Psi^{NN} }} = \sum_{\bm{\sigma}}  \frac{\lvert \Psi^{NN}(\bm{\sigma}) \rvert^2}{\braket{\Psi^{NN} | \Psi^{NN} }} O_\text{local}(\bm{\sigma}),
\end{equation}
where $O_\text{local}(\bm{\sigma})=\sum_{\bm{\sigma}'} \braket{\bm{\sigma}| \hat{O} | \bm{\sigma}'} \Psi^{NN}(\bm{\sigma}) / \Psi^{NN}(\bm{\sigma}')  $. 
This expectation value can be estimated using Markov Chain Monte Carlo (MCMC) sampling from the normalized probability $P_{\Psi^{NN}}(\bm{\sigma})$.
One can then minimize the energy expectation value by stochastic gradient descent (SGD).
This method is also known as variational Monte Carlo (VMC)~\cite{mcmillan1965ground,foulkes2001quantum,yokoyama1987variational,sorella2005wave} and is often combined with the second-order optimization method, stochastic reconfiguration~\cite{sorella2001generalized,PhysRevB.71.241103,neuscamman2012optimizing}, a.k.a. imaginary time-dependent variational principle (iTDVP) or natural gradient descent~\cite{amari1998natural}.
While being a versatile method, the NQS-based method has potential drawbacks including the intractable normalization and the long auto-correlation time in MCMC, which can lead to difficulties in training large deep neural networks.

\subsection{ Neural Autoregressive Quantum States \label{subsec: NAQS}}

A proposed solution to resolve the potential drawbacks mentioned above is the neural autoregressive quantum states (NAQS)~\cite{sharir2020deep,hibat2020recurrent,luo2020probabilistic,luo2021gauge}.
NAQS are NQS with causal restrictions in the connectivity of the networks.
Concretely, NAQS approximate the conditional probability amplitudes of a given ordering of sites $\{1,\ldots,N\}$ by a neural network,
\begin{align}
    &\Psi_{\sigma_1 \sigma_2 \cdots \sigma_N} 
    = \Psi_{\sigma_1} \Psi_{\sigma_2|\sigma_1} \cdots \Psi_{\sigma_N|\sigma_1 \cdots \sigma_{N-1}} \\
    &\qquad \approx f_\text{NN}^{[1]}(\sigma_1) f_\text{NN}^{[2]}(\sigma_2|\sigma_1) \cdots f_\text{NN}^{[N]}(\sigma_N|\sigma_1 \cdots \sigma_{N-1}) 
\end{align}
where the network has $N$ outputs.
Each output represents a conditional probability amplitude on each site $\Psi_j\coloneqq \Psi_{\sigma_j | \sigma_{i<j}} $.
The fact that $\Psi_j$ can only depend on $\sigma_{i<j}$ leads to a causal structure.
Since every single site conditional probability amplitude $\Psi_j$ can be easily normalized, the multiplication of them gives a normalized variational quantum state (see Fig.~\ref{fig: ansatz}).

As a subclass of NQS, NAQS share the same properties of NQS where one can efficiently evaluate the probability amplitudes and the gradients.
Furthermore, a NAQS represents an autoregressive model.
An autoregressive model can be sampled by the ancestral sampling algorithm which allows sampling directly from the probability $P_{\Psi^{NN}}(\bm{\sigma}) = \lvert \Psi^{NN}(\bm{\sigma}) \rvert^2 $ (see Appendix~\ref{appendix: NAQS} for further details).
As a result, quantum states represented by a deep neural network with millions of parameters can be trained efficiently using the SGD algorithm for studying ground states~\cite{sharir2020deep}.

\subsection{ Supervised learning for Quantum State Approximation \label{subsec: supervised_learning}}

We now describe the variational problem of approximating complex-valued wavefunctions in the supervised learning setting and propose a new cost function for normalized variational wavefunctions, e.g. NAQS, to jointly learn the probabilities and phases.

Given a target wave function $| \Phi \rangle$, we would like to find the optimal NAQS $| \Psi^\text{NN} \rangle$ approximating the target wave function by minimizing a cost function $\mathcal{R}(\Phi, \Psi^\text{NN})$.
In the following, we consider the setup that we are given the samples $\{ \bm{\sigma}, \Phi(\bm{\sigma}) \}$ according to the probability distribution $P_{\Phi} (\bm{\sigma} ) =  \lvert \Phi(\bm{\sigma}) \rvert^2$ instead of the full wave function.
It is a practical setup for the exponentially large Hilbert space.
In the supervised learning setup, a loss function $\mathcal{L}$ measures the difference between the data points $\braket{\bm{\sigma}|\Phi}, \braket{\bm{\sigma}|\Psi^\text{NN}}$.
A cost function $\mathcal{R}$ is the expectation value of the loss function $\mathcal{L}$ with respect to the probability $P_{\Phi} (\bm{\sigma} )$ and is therefore a measure of the difference between the quantum states $\ket{\Phi}, \ket{\Psi^\text{NN}}$.
As a result, the optimization problem can be solved using the SGD method, where the stochastic gradient is computed from the sampled cost function $\hat{R}$ (see Appendix~\ref{appendix: cost} for a detailed discussion).

Here we present a joint learning scheme for complex-valued wave function where the cost function $\mathcal{R}_\text{joint}$ is the sum of the so-called Kullback–Leibler (KL) divergence of the probabilities $\mathcal{R}_\text{KL}$ and the weighted $\mathcal{L}_2$ distance of the phases $\mathcal{R}_{\theta}$,
\begin{equation}
    \mathcal{R}_\text{joint} = \mathcal{R}_\text{KL} + \mathcal{R}_{\theta}.
\end{equation}

In order to learn the magnitude of the probability amplitude $\text{Re}\left[ \log \Psi^\text{NN}(\bm{\sigma}) \right ]$, we consider the forward KL-divergence $\mathcal{R}_\text{KL}$, which is defined as
\begin{align}
\begin{split}
\label{eq:cost_kl}
\mathcal{R}_\text{KL}
&= \sum_{i}^{D_\mathcal{H}} P_{\Phi}(\bm{\sigma}_i) \log{\left(\frac{P_{\Phi}(\bm{\sigma}_i)}{P_{\Psi^\text{NN}}(\bm{\sigma}_i)} \right)} \\
&\approx \frac{ \sum_{\bm{\sigma}_i\sim |\Phi(\bm{\sigma}_i)|^2}  2\text{Re}\left[  \log \Phi(\bm{\sigma}_i) - \log \Psi^\text{NN}(\bm{\sigma}_i) \right ]. }{N_\text{samples}}
\end{split}
\end{align}
It is a common cost function for learning a probability distribution when one can sample from the exact distribution, which is equivalent to maximal likelihood learning.

For learning the phase of the probability amplitude $\text{Im}\left[ \log \Psi^\text{NN}(\bm{\sigma}) \right ]$, we consider the weighted $\mathcal{L}_2$ distance of the phases embedded on a unit circle,
\begin{align}
\begin{split}
\label{eq:cost_theta}
\mathcal{R}_\theta
&=  \sum_{i}^N |\Phi(\bm{\sigma}_i)|^2 \ \text{dist}(\text{Im}\left[ \log \Phi(\bm{\sigma}_i) \right], \text{Im}\left[ \log \Psi^\text{NN}(\bm{\sigma}_i) \right])\\
&=  \sum_{i}^N |\Phi(\bm{\sigma}_i)|^2 \ \text{dist}(\theta^\Phi(\bm{\sigma}_i),\ \theta^{\Psi^\text{NN}}(\bm{\sigma}_i) )\\
&\approx \frac{1}{N_\text{samples}} \sum_{\bm{\sigma}_i\sim |\Phi(\bm{\sigma}_i)|^2} \text{dist}(\theta^\Phi(\bm{\sigma}_i),\ \theta^{\Psi^\text{NN}}(\bm{\sigma}_i) )  \\
\end{split},
\end{align}
with
\begin{equation*}
     \text{dist}(\theta_1,\ \theta_2) =   %
       \big( \cos\left( \theta_1  \right) - \cos\left( \theta_2 \right) \big)^2 +  \big( \sin\left( \theta_1 \right) - \sin\left( \theta_2 \right) \big)^2 
       .
\end{equation*}

The stochastic gradient can be computed from the sampling estimate of the cost functions in Eq.~\eqref{eq:cost_kl} and Eq.~\eqref{eq:cost_theta}. 
During the optimization procedure, the magnitudes and phases are learned at the same time.
This cost function is only valid for normalized wavefunction because the KL-divergence is unbounded from below if $P_{\Psi^\text{NN}}(\bm{\sigma}_i)$ is an unnormalized probability distribution. 
As a result, we consider NAQS which ensures the normalization without additional cost.

The advantage of choosing the cost function $\mathcal{R}_\text{joint}$ is that the resulting stochastic gradient is unbiased and with low variance.
The optimization problem can be solved by SGD with a small mini-batch size or even a single data input at a time.
It is important to have this property for efficiently solving the variational state approximation problem with a large neural network and when the Hilbert space is large.

Another cost function, which also can be rewritten as an expectation value and gives a physical interpretation, is the real part of the negative overlap,
\begin{align}
\begin{split}
\label{eq:cost_overlap}
\mathcal{R}_\text{neg. overlap}
&= - \text{Re} \left [ \sum_{i}^{D_\mathcal{H}} {\Phi}^*(\bm{\sigma}_i)\Psi^\text{NN}(\bm{\sigma}_i) \right ]\\
&= -\sum_{i}^{D_\mathcal{H}}  P_{\Phi}(\bm{\sigma}_i) \text{Re}\left[  \frac{\Psi^\text{NN}(\bm{\sigma}_i)}{\Phi(\bm{\sigma}_i)}   \right]
\end{split}
\end{align}
We observed that it has similar performance as the joint cost function. In Appendix~\ref{appendix: cost}, we show the two cost functions are indeed equivalent to first order in error.
In principle, one should include a constant term $c_\text{ratio}$ in the equation $\mathcal{R}_\text{joint} = \mathcal{R}_\text{KL} + c_\text{ratio} \mathcal{R}_{\theta}$, where the $c_\text{ratio}$ is a hyper-parameter to be tuned controlling the ratio between the two cost functions.
We consider $c_\text{ratio}=1$ because the equivalence to $\mathcal{R}_\text{neg. overlap}$ in the small error limit.

In passing we mention, that several different proposals for the cost functions have been discussed in the literature. This includes the $\mathcal{L}_2$ distance of the probability amplitudes~\cite{cai2018approximating,kochkov2018variational}, the weighted $\mathcal{L}_2$ distance of log of the probability amplitudes~\cite{westerhout2020generalization}, and negative log fidelity~\cite{jonsson2018neural}.
We discuss the detail of the difference between different cost functions in Appendix~\ref{appendix: cost}

Note that the variational algorithm to approximate quantum states using neural networks has found applications in imaginary-time evolution~\cite{kochkov2018variational}, real-time evolution~\cite{gutierrez2019real}, simulating open quantum system~\cite{luo2020probabilistic}, classical simulation of quantum computation~\cite{jonsson2018neural} and the study of the expressivity and generalization properties~\cite{cai2018approximating,carleo2019netket,borin2020approximating,westerhout2020generalization}.

\section{Numerical Results  \label{sec: te}}

To test the variational power of different networks, we consider far-from-equilibrium dynamics resulting from global quenches for a paradigmatic 1D model.
The fast (ballistic) growth of entanglement prohibits the classical simulation based on MPS at long times.
Recent work with NQS suggests that NQS could reach time scales comparable with or exceeding those of the state-of-the-art tensor network methods~\cite{schmitt2020quantum}.
Here, we consider a similar setup with a $x$-polarized initial product state $|\psi_0\rangle = \prod_i |\rightarrow \rangle_i$.
We then evolve the state using the Hamiltonian of the quantum Ising model with transverse $(g)$, longitudinal $(h)$ fields and an interaction term $(k)$:
\begin{equation}\label{eq: Hamiltonian}
    \hat{H} = - J \left[\sum_{j=1}^{N-1} 
    \left( 
    \hat{\sigma}^z_j \hat{\sigma}^z_{j+1} + 
    k\hat{\sigma}^x_j \hat{\sigma}^x_{j+1} \right) 
    + \sum_{j=1}^{N} 
    \left (
    g \hat{\sigma}^x_j +  
    h \hat{\sigma}^z_j
    \right)
    \right].
\end{equation}
We consider two cases that exhibit different growth of entanglement: a weak quench to $g=3$, $h=0.1$, $k=0$ with a slow growth of entanglement and a strong quench near the critical point $g=1$, $h=0$, $k=0.25$ with fast growth. 
Note that the model is non-integrable for both parameter sets.
In the following, we consider open boundary condition (OBC) and provide additional data for periodic boundary condition (PBC) in Appendix~\ref{appendix: data}.

\subsection{Approximation with MPS \label{subsec: te_mps}}

\begin{figure}[t]
\centering
\includegraphics[width =\columnwidth]{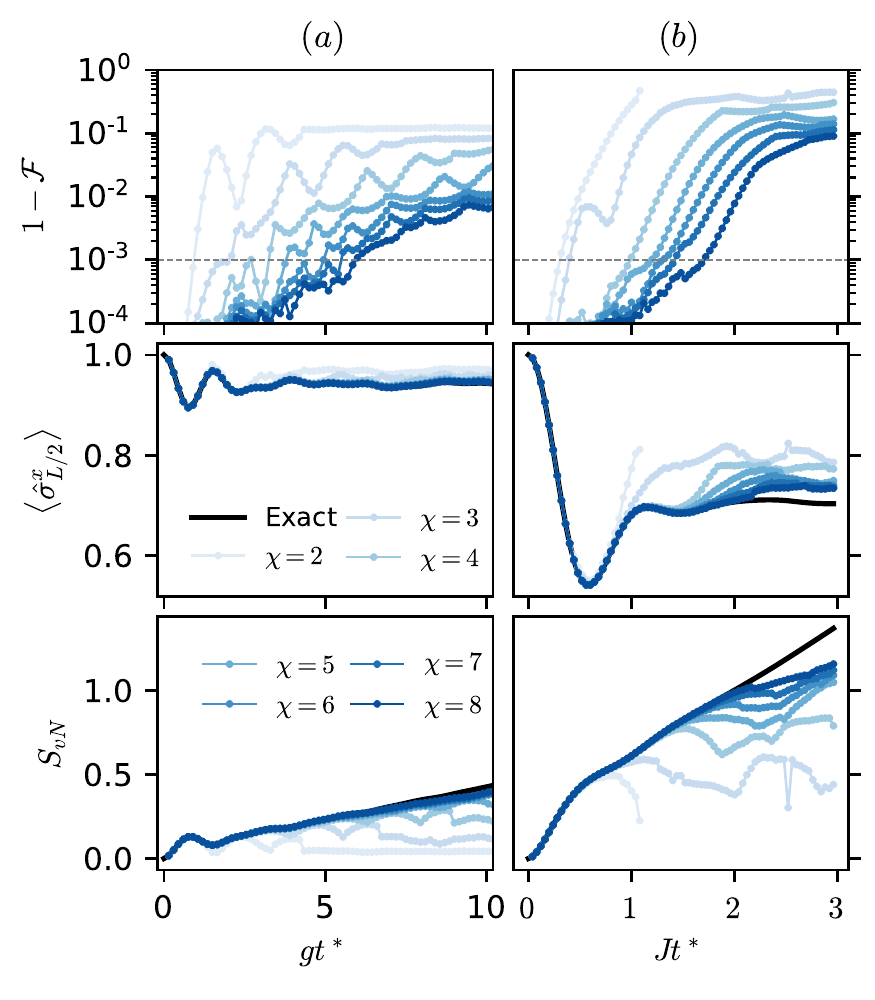}
\caption{MPS with bond dimension $\chi$ approximating the exact time-evolved states following a quantum quench from an initial paramagnetic product state.
The MPS are obtained by supervised learning with SGD.
The quenched Hamiltonian is given in Eq.~\eqref{eq: Hamiltonian} for a chain of length $N=20$ with (a) quantum Ising model in the paramagnetic phase with weak longitudinal field ($g=3$, $h=0.1$, $k=0$) and (b) close to critical point with interacting term ($g=1$, $h=0$, $k=0.25$).
\label{fig: MPS_data}}
\end{figure}

We first variationally approximate the exact time-evolved state using the MPS ansatz with small bond dimension $\chi$ and reproduce the well known result that the number of parameters scales exponentially in time.
In particular, we perform the exact simulation of time evolution following the quench to obtain the target states $|\Phi_\text{exact}(t)\rangle$ at certain discrete times using ED.
We then find the MPS with given bond dimension $\chi$ approximating the target state at each time $t$ by SGD over the cost function $\mathcal{R}_\text{joint}$.
Note that the MPS found by supervised learning with SGD method is close to the optimal result obtained by iteratively minimizing the $\mathcal{L}_2$ distance of the quantum states~\cite{schollwock2011density}.
We include this additional result in Appendix~\ref{appendix: data} for comparison. The details of the parameters used in the optimization are shown in Appendix~\ref{appendix: training_setup}.

In Fig.~\ref{fig: MPS_data} we show the error in fidelity after the quantum state approximation for the two parameter sets of the Ising model. 
We also show the expectation value $\langle \hat{\sigma}_x \rangle$ in the middle of the chain and the half-chain von Neumann entanglement entropy $S_{vN}$.
For a fixed bond dimension $\chi$, an MPS can accurately approximate the state under time evolution to a certain timescale after which the error in fidelity grows exponentially to a saturation value.
This timescale coincides with the deviation in the local observable $\langle \hat{\sigma}_x \rangle$ and the saturation of entanglement.

To quantitatively access this time scale, we determine the reachable time $t^*$ by the time when the error in fidelity exceeds a threshold value $1-\mathcal{F}=10^{-3}$.
This is indicated by the grey dashed lines in Fig.~\ref{fig: MPS_data}.
We then plot the number of parameters of the MPS as a function of the time $t^*$ that can be reached in Fig.~\ref{fig: para_time_scaling}.
The data shows a clear exponential growth of the number of parameters as a function of  $t^*$.
This is directly related to the linear growth of the entanglement $S$ entropy~\cite{PhysRevLett.111.127205}, which yields an exponentially growing bond dimension $\chi\sim e^{S}$ to correctly capture the underlying states~\cite{PhysRevB.73.094423,eisert2013entanglement}.

Note that the exponential scaling of the number of parameters with $t^*$ is not affected by the threshold value chosen as long as the threshold value for the error in fidelity $1-\mathcal{F}$ is chosen below the value where the curves flatten due to finite size effects.

\subsection{Approximation with shallow NAQS \label{subsec: te_naqs}}

We now investigate the variational power of NAQS in a similar setup as before.
Namely, given the exact time-evolved state $\ket{\Phi(t)}$ at the selected time $t$, we perform supervised learning to approximate the state variationally with NAQS.
As a first example, we consider a one-hidden layer convolutional (CNN-1) NAQS. 
The ratio of the number of hidden neurons to the system size $\alpha=N_\text{hidden}/N$ controls the width of the network.
For CNNs, $\alpha$ is the number of channels.
Increasing the width of the network increase the number of parameters and the variational power of the network.
We optimize the cost function described in Section~{\ref{subsec: supervised_learning}} using SGD with the Adam~{\cite{kingma2014adam}} optimizer. We begin with a batch size of $512$ samples and a learning rate of $10^{-3}$ which is then decreased to $10^{-4}$ once the sampled cost converges.
Technical details are described in Appendix~{\ref{appendix: NN}}.

\begin{figure}[t!]
\centering
\includegraphics[width=\columnwidth]{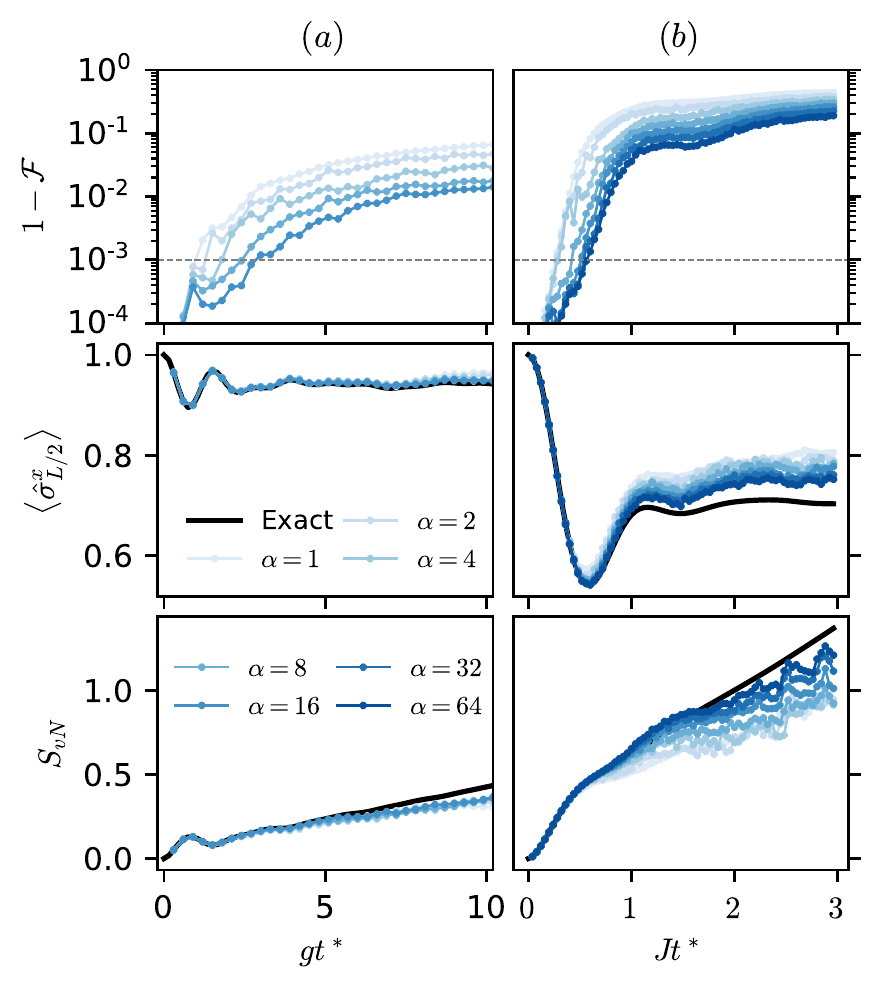}
\caption{One-hidden layer convolutional NAQS of different width $\alpha$ approximating the exact time-evolved states following a quantum quench from an initial paramagnetic product state.
The quenched Hamiltonian is given in Eq.~\eqref{eq: Hamiltonian} for a chain of length $N=20$ with (a) quantum Ising model in the paramagnetic phase with weak longitudinal field ($g=3$, $h=0.1$, $k=0$) and (b) close to critical point with interacting term ($g=1$, $h=0$, $k=0.25$).
\label{fig: NN_data}}
\end{figure}

In Fig.~\ref{fig: NN_data}, we show the results obtained for the quantum Ising model described in Eq.~(\ref{eq: Hamiltonian}).
We observe for a fixed network that the accuracy of the approximation decreases with time.
The error increases exponentially and saturates at a size dependent final value as in the case of MPS.
Moreover, we find that the accuracy improves when increasing the width $\alpha$ of the networks.
Identically to the procedure for MPS, we determine the reachable time $t^*$ by the time when the error in fidelity exceeds the threshold value $1-\mathcal{F} = 10^{-3}$ indicated by the grey dashed lines.
We notice that the entanglement entropy also deviates from the exact value but does not saturate even at long times after the accessible time $t^*$.
This result is plausible since NAQS can indeed represent volume law states.

\begin{figure}[t!]
\centering
\includegraphics[width =\columnwidth]{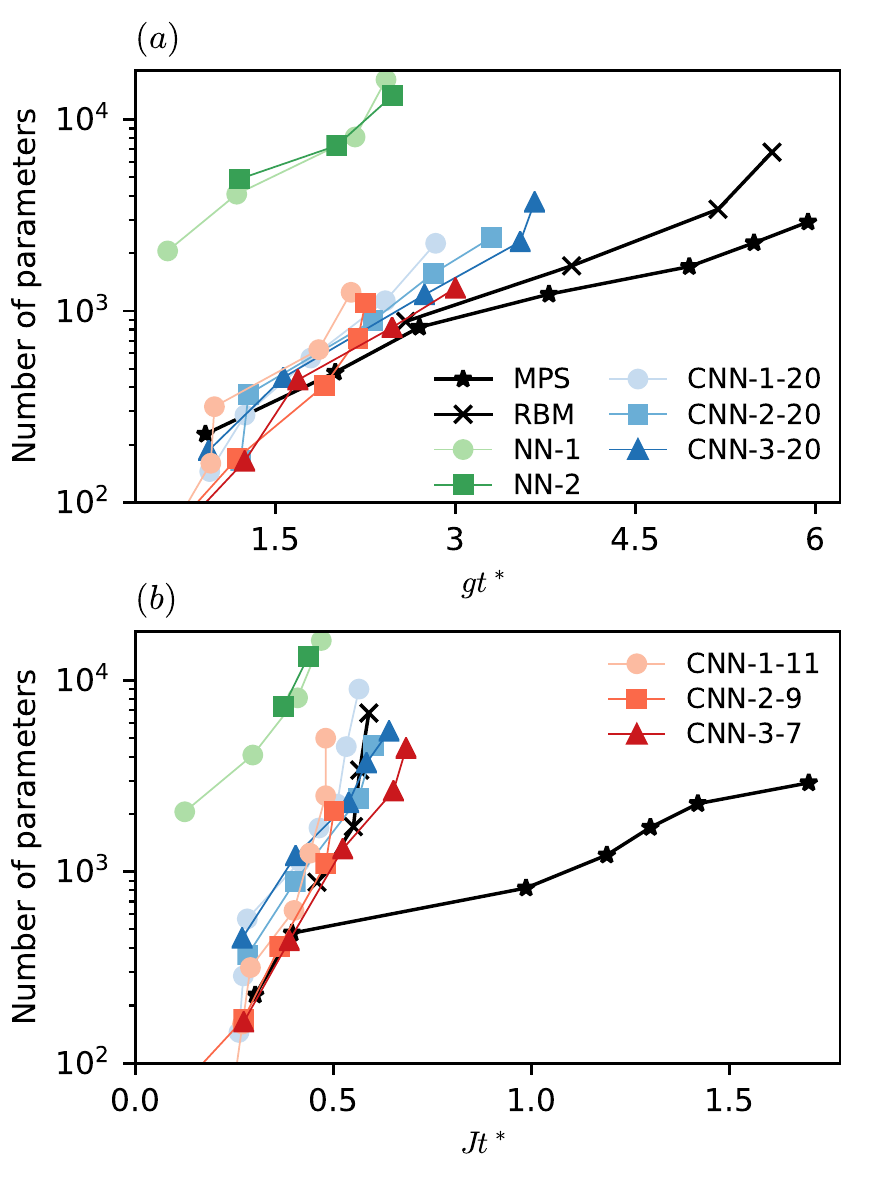}
\caption{ 
Effect of increasing network width: The number of parameters required to reach time $t^*$ for following a quantum quench for different approximation schemes.
The quenched Hamiltonian is given in Eq.~\eqref{eq: Hamiltonian} for a chain of length $N=20$ with (a) quantum Ising model in paramagnetic phase with weak longitudinal field ($g=3$, $h=0.1$, $k=0$) and (b) close to critical point with interacting term ($g=1$, $h=0$, $k=0.25$).
The black line with star (cross) marker shows the results for MPS (RBM).
For NAQS, we denote each combination as (network type)-(number of hidden layers)-(filter size).
The circles, squares, triangles represent 1,2,3-layer networks respectively.
The gradient colors describe same networks of different number of hidden layers.
NNs (Blue); CNNs with global connectivity (Green-Blue); CNNs with local connectivity (Red).
}\label{fig: para_time_scaling}
\end{figure}

In Fig.~\ref{fig: para_time_scaling}, we plot the number of parameters against this reachable time $t^*$ with various widths $\alpha$.
The result suggests an exponential scaling of the number of parameters with the reachable time $t^*$.
This is identical to the scaling behavior of MPS but the growth constant is in fact larger.

A big advantage of NAQS is the flexibility in the design of network architectures.
Essentially, NAQS are broad classes of wavefunctions based on different network architectures.
With a fixed network architecture, two common factors to control the model complexity are the network width and network depth~\cite{he2016deep,zagoruyko2016wide,xie2017aggregated}.
An important question is how these choices affect the network expressivity, hence the quality of the approximation.

To understand the effect of the network width on the expressivity, we focus on shallow NNs with at most 3-hidden layers and consider three types of network architectures: (i) fully-connected neural networks (NNs), (ii) CNNs with a large filter size equaling to the system size, and (iii) CNNs with small filter sizes but have receptive fields (causal cone) covering the full system.
For simplicity, we fix the same network width $\alpha$ over the hidden layers.
For a given combination of the network type and the number of hidden layers, we repeat the procedure of state approximation described above with a increasing network width $\alpha$.
We plot the number of parameters of the network against the obtained accessible times $t^*$ in Fig.~\ref{fig: para_time_scaling}.
We denote each combination as (network type)-(number of hidden layers) and with (filter size) if a CNN is considered.

From the data, we observe consistent exponential scaling of the number of parameters with the reachable time $t^*$ by increasing the width $\alpha$ for all network architectures and depths considered.
Changing the network architecture results in a constant shift and a slight change in the growth rate in the scaling.
There is a significant difference between NNs and CNNs, where we see an order of magnitude improvement likely due to the sharing of parameters.
The difference between CNNs with small and large filter sizes is, however, not obvious.
While here we focus on shallow neural networks, the data still shows a slight improvement in the growth rate when increasing the network depth.
Notably, the scaling of NNs deviates more from that of MPS for parameter set (b). 
It is different from the naive expectation that MPS would perform less efficiently when the entanglement grows rapidly.

We would like to point out that the scaling does not depend on boundary conditions---the same exponential scaling is observed for systems with PBC (see Appendix~\ref{appendix: data}).
Due to the causality requirement of the NAQS, we have zero-padding instead of periodic padding for CNNs.
This implies that the wavefunction is not translational invariant; hence, it can be applied to both OBC and PBC systems.
While we found a significant constant shift in scaling between CNNs and NNs due to parameter sharing, we expect an even more significant shift for other types of NQS where one enforces exact translational symmetry.

\subsection{Approximation with deep NAQS \label{subsec: te_naqs_deep}}

To study the effect of network depth on expressivity, we consider two types of deep network architectures: (i) CNNs with gated activation unit and residual connection (Gated-CNNs) (ii) WaveNets~\cite{oord2016wavenet}.
Both networks consist of modular building blocks.
One can systematically increase the network depth by stacking more blocks.
The building block of Gated-CNNs includes causal convolution following a gated activation unit and a residual connection~\cite{he2016deep}.
The building block of WaveNets replaces the standard causal convolution with dilated causal convolution, which allows the size of the receptive field (causal cone) to grow exponentially with the network depth, and it includes an additional parameterized skip connection.
We show the building block and the network architecture in Fig~\ref{fig: wavenet} and describe the detail of the network architectures in Appendix~\ref{appendix: NN}.
For a fixed network architecture and width $\alpha$, we increase the network depth to increase the number of parameters and repeat the procedure of state approximation.
We vary the number of blocks between $6$ to $14$ for WaveNets and $10$ to $20$ for Gated-CNNs.

\begin{figure}[t!]
\centering
\includegraphics[width=\columnwidth]{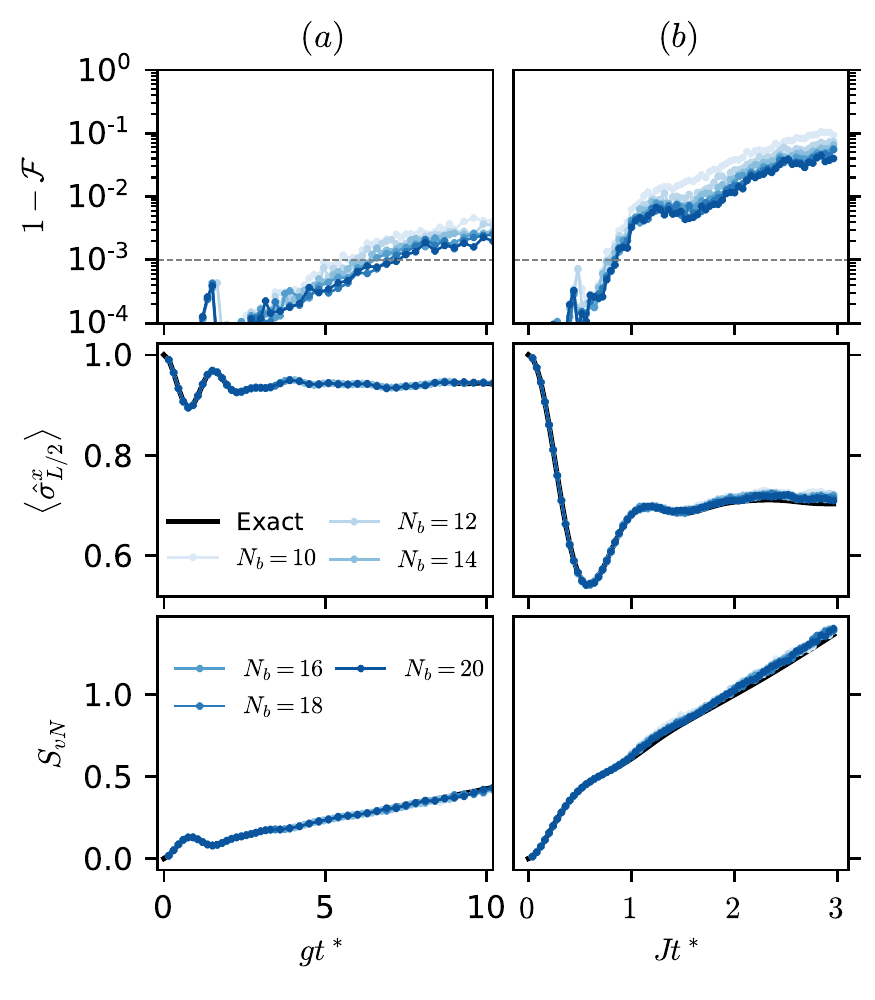}
\caption{Gated-CNNs NAQS of different depth, i.e. number of blocks $N_b$, approximating the exact time-evolved states following a quantum quench from an initial paramagnetic product state.
The network is of width $\alpha=12$.
The quenched Hamiltonian is given in Eq.~\eqref{eq: Hamiltonian} for a chain of length $N=20$ with (a) quantum Ising model in the paramagnetic phase with weak longitudinal field ($g=3$, $h=0.1$, $k=0$) and (b) close to critical point with interacting term ($g=1$, $h=0$, $k=0.25$).
\label{fig: deep_NN_data}}
\end{figure}

In Fig.~\ref{fig: deep_NN_data}, we show the results for Gated-CNNs of width $\alpha=12$.
We find a similar result as in Fig.~\ref{fig: NN_data}.
With a larger network, it is now obvious that the entanglement entropy deviates but does not saturate long after the accessible time.
We even observe slightly larger entanglement entropy at the deviation (See also data for WaveNets in Appendix~\ref{appendix: data}).

In Fig.~\ref{fig: deep_para_time_scaling}, we plot the number of parameters of networks with different depths against the accessible times $t^*$ and denote each combination as (network type)-(network width $\alpha$).
We observe again consistent exponential scaling of the number of parameters with the reachable time $t^*$ by increasing the network depth for the two networks and widths considered.
Similarly, we see a small constant shift for different network width favoring ``narrower" networks.
More importantly, data for both network architectures roughly fall on the same line despite the drastically different design choice of using dilated or normal convolution.
Compared with the data from shallow CNNs where we vary the network width, all data again fall roughly on the same line.
This suggests that increasing the number of parameters by increasing network width has the same effect on expressivity as increasing the network depth for the cases considered.
Even more surprisingly, such results are independent of the network architectures of whether one considers global or local, dilated or normal convolution, with or without skip and residual connections.
Note that, however, the optimization problem may be strongly affected by the choice of different network architectures.
For example, a deep neural network without shortcut connections may be hard to optimize.

\begin{figure}[t!]
\centering
\includegraphics[width =\columnwidth]{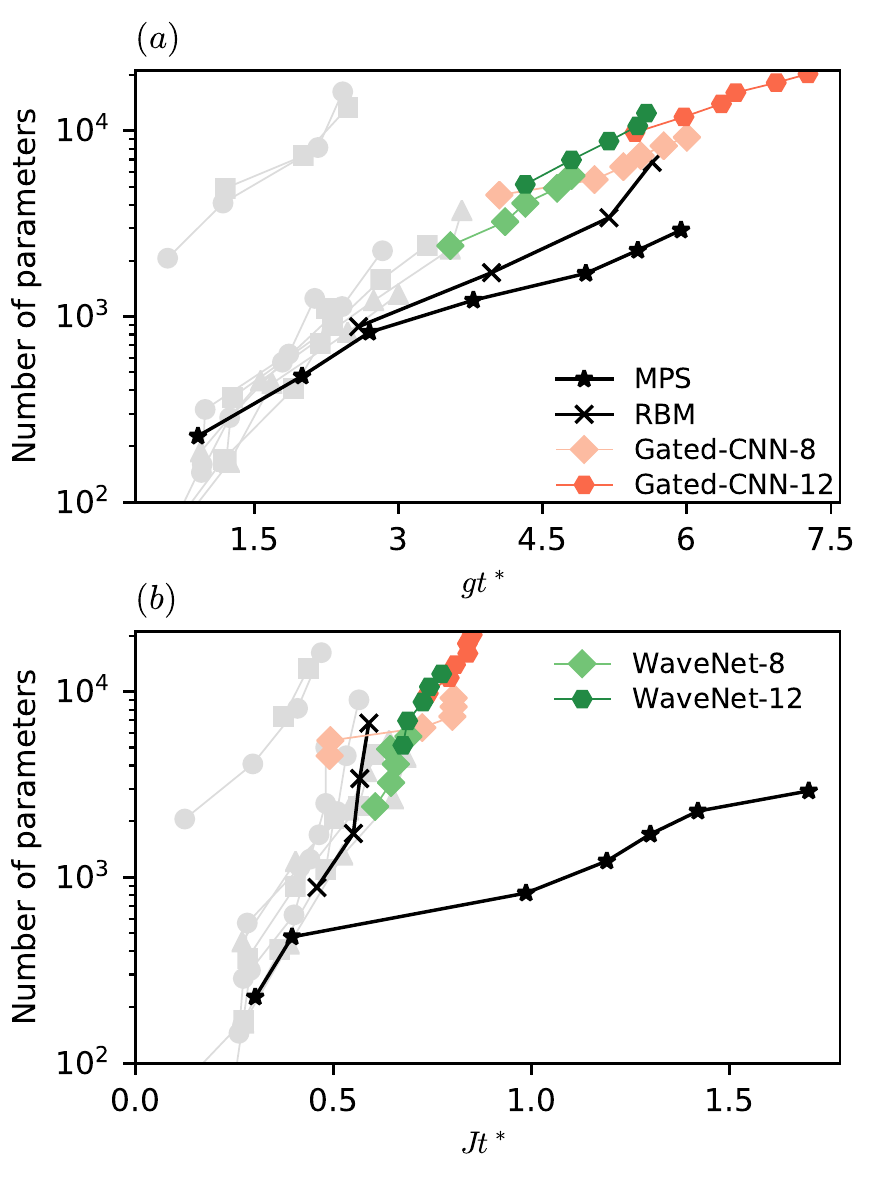}
\caption{
Effect of increasing network depth:
The number of parameters required to reach time $t^*$ for following a quantum quench for following a quantum quench for different approximation schemes.
The quenched Hamiltonian is given in Eq.~\eqref{eq: Hamiltonian} for a chain of length $N=20$ with (a) quantum Ising model in paramagnetic phase with weak longitudinal field ($g=3$, $h=0.1$, $k=0$) and (b) close to critical point with interacting term ($g=1$, $h=0$, $k=0.25$).
The black line with star (cross) marker shows the results for MPS (RBM).
For NAQS, we denote each combination as (network type)-(network width $\alpha$).
The diamonds and hexagons represent network of width $\alpha=8,12$ respectively.
The gradient colors describe same networks of different network width.
Gated-CNNs (Red); WaveNets (Green); we keep the shallow NAQS data (light grey) from Fig.~\ref{fig: para_time_scaling} for comparison.
}\label{fig: deep_para_time_scaling}
\end{figure}

We further note that there are two main differences in the network structure considered here and in previous works~\mbox{\cite{carleo2017solving,PhysRevB.98.024311,schmitt2020quantum,gutierrez2019real}}.
The first difference is the autoregressive property.
It is unknown how the autoregressive constraint affects the expressivity of the neural networks.
The second difference is the activation function.
In Ref.~\mbox{\onlinecite{schmitt2020quantum,gutierrez2019real}}, the choice of polynomial activation with complex-valued weight is crucial to the result.
Notice that such networks~\mbox{\cite{voigtlaender2020universal}} do not have the universal approximation property.
In contrast, we consider real-valued weights with rectified linear unit (ReLU)~\mbox{\cite{nair2010rectified}}, tanh, and sigmoid activation functions.
For shallow NNs, we observe that changing activation function does not affect the result significantly.
The difference in the setup calls into question whether the good result before~\mbox{\cite{schmitt2020quantum}} is not a general property of neural networks.

\subsection{ Approximation with RBMQS \label{sec: RBMQS}}

As the last example, we study the expressivity of the restricted Boltzmann machine quantum states (RBMQS)~\cite{carleo2017solving} with increasing width.
One prominent difference between RBMQS and NAQS is that RBMQS belongs to general (non-autoregressive) neural network, hence is not normalized.
The normalization constant is in general intractable.
Nevertheless, it is possible to estimate the fidelity to another wavefunction by stochastic sampling.
The gradient of such stochastic estimate gives an biased stochastic gradient, which works well in practice for supervised learning with stochastic gradient methods.

We repeat the similar benchmark of state approximation on networks of increasing width $\alpha=N_\text{hidden}/N=\{1,2,4,8\}$.
We minimize the negative fidelity between the RBMQS and the target state by SGD.
We confirm our implementation yields similar result to the NetKet~\cite{carleo2019netket} implementation using stochastic reconfiguration (SR) method.
For more detail, see the discussion of the cost function in Appendix~\ref{appendix: cost} and the additional data for consistency check in Appendix~\ref{appendix: data}.

The result for state approximation using RBMQS is shown in Fig.~\ref{fig: RBM} in Appendix~\ref{appendix: data}, which resembles the results of NAQS shown in Fig.~\ref{fig: NN_data} and Fig.~\ref{fig: deep_NN_data}.
We plot the reachable time $t^*$ in Fig.~{\ref{fig: para_time_scaling}} and Fig.~{\ref{fig: deep_para_time_scaling}}.
We observe again consistent exponential scaling of the number of parameters with the reachable time $t^*$ by increasing the network width.

We would like to point out our observation in the reachable time is consistent with  previous works~{\cite{carleo2017solving,PhysRevB.98.024311}} which show similar difficulties around $Jt=0.5$ for representing states quenched by Hamiltonian near the critical point using RBM-based NQS.
It is also observed in~{\cite{gutierrez2019real}} that a lack of expressibility from the RBM-based NQS leads to a rapid increase in the error in fidelity in a short time scale for a quench across critical point.

Lastly, similar scaling plots as in Fig.~{\ref{fig: para_time_scaling}} and Fig.~{\ref{fig: deep_para_time_scaling}} could also be obtained in a more realistic setup when exact data is not known.
For MPS, one could perform the time-evolving block decimation (TEBD) algorithm~{\cite{vidal2003efficient,vidal2004efficient}}, and only increase the bond dimension $\chi$ when the approximation (truncation) error reaches the threshold value $10^{-3}$.
For NQS, one could perform a time evolution algorithm based on supervised learning~{\cite{gutierrez2019real,luo2020probabilistic}} and increase the size of the neural network when the approximation error reaches the threshold value $10^{-3}$.
If we plot the number of parameters of the variational wavefunctions to the time steps, we could obtain a similar scaling plot.

\section{ Conclusion \label{sec: conclusion}}

We provide numerical evidence that the required number of parameters to represent quantum states following a global quench by neural networks grows exponentially in time.
Thus, despite the ability to represent highly-entangled states, the neural network quantum states (NQS), including neural autoregressive quantum states (NAQS) and restricted Boltzmann machine quantum states (RBMQS), considered in this work cannot represent time-evolved states efficiently.
An important remaining question is to understand this scaling.
The scaling is agnostic to the design of network architectures we considered but is only affected by the difference in architecture related to symmetry, i.e. convolution.
The scaling of NQS resembles that of MPS. However, in the one-dimensional models considered, NQS shows a larger growth rate of parameters in time while its entanglement entropy saturates slower.
Explaining this difference and understanding the limitation of NQS may require a new measure for the complexity of states different from the entanglement entropy.

The proposed cost function for joint learning magnitudes and phases with NAQS ensures unbiased and low-variance stochastic gradient.
It could be applied to the real-time evolution for pure states and mix states~\cite{gutierrez2019real,luo2020probabilistic} and the imaginary-time evolution~\cite{kochkov2018variational}.
We expect it to speed up the state approximation at each time step and to be suitable for learning with deep neural networks.
Monitoring the training and convergence of the two costs $\mathcal{R}_\text{KL}$ and $\mathcal{R}_{\theta}$ gives information about the hardness of learning probability distributions and the phases.
We observe similar difficulty in learning both the magnitudes and the phases for time-evolved states.
This may provide additional insight for learning frustrated ground states, which previous works~\cite{szabo2020neural,westerhout2020generalization} show the main difficulty comes from learning the phases.

Finally, our findings do not invalidate the simulation for time evolution with NQS in general but rather provides insight on the choice of networks to reach the desired time scale.
It also suggests the importance to consider dynamically increasing the network size for the time evolution algorithm based on supervised learning.
It is still important to search for different network architectures~\cite{he2016deep,huang2017densely,vaswani2017attention,oord2016conditional,salimans2017pixelcnn++,oord2016wavenet}, which might be more efficient in optimization or better in the scaling for different Hamiltonians.
Network architectures incorporating the symmetry of the states have the potential to give better scaling.
Moreover, the approach based on neural networks is less affected by the spatial dimension of the systems than that based on tensor networks.
This suggests the potential advantage of NQS over tensor network states in two and higher dimensions.

While understanding the expressivity and scaling of NQS is the essential step for designing practical algorithms, the recent work~\mbox{\cite{hofmann2021role}} studies the numerical instability in obtaining the NQS using time-dependent variational Monte Carlo when the time-evolved states can be expressed as NQS accurately.
The article~\mbox{\cite{hofmann2021role}} and this work complement each other and provide insights for simulating time evolution using NQS.
The code implementation of this work is based on TensorFlow~\mbox{\cite{tensorflow2015-whitepaper}} and is available at~\mbox{\cite{GitHub2021}}.

\begin{acknowledgments}
	\noindent S.L. would like to thank Markus Schmitt, Rohit Dilip, Adam Smith, and Jin-Guo Liu for helpful discussion.
	This work is supported by the European Research Council (ERC) under the European Union's Horizon 2020 research and innovation program (grant agreement No. 771537). 
	S.L. and F.P. are supported by the Deutsche Forschungsgemeinschaft (project number 107745057, TRR 80). 
	F.P. acknowledges the support of the Deutsche Forschungsgemeinschaft (DFG, German Research Foundation) under Germany's Excellence Strategy EXC-2111-390814868. 
\end{acknowledgments}

\appendix

\section{Network Architectures \label{appendix: NN}}

Feed-forward neural networks are functions consisting of alternating affine maps and elementwise non-linear functions.
The input of the network is the one-hot encoding of the computational basis. That is
\begin{align}
    &X : \bm{\sigma}\in\{1,2,\ldots,M\}^N \longrightarrow X(\bm{\sigma})\in \mathbb{R}^{M\times N}\\
    &X(\bm{\sigma})_{i,j} = \begin{cases}
1,\qquad \text{if}\ i=\sigma_j \\
0,\qquad  \text{else}
\end{cases}
\end{align}
Fully-connected neural networks (NNs) are defined by a recursion relation at each layer
\begin{equation}
    \textbf{z}^{(l)} = g({W}^{(l)}\textbf{z}^{(l-1)} + \textbf{b}^{(l)}),
\end{equation}
where the weight matrix $W^{(l)}$ and bias vector $\textbf{b}^{(l)}$ at each layer $(l)$ together are the variational parameters.
The $g$ is the elementwise non-linear function, a.k.a activation function.
At the first layer, the input is first flattened as a vector, i.e. $z^{(0)} = \text{Vec}(X(\bm{\sigma}))$.

We consider NNs with one- and two-hidden layers and denote them as NN-1 and NN-2.
The number of neurons are $(N,\alpha N, 4N)$ for NN-1 and $(N,\alpha N, \alpha N, 4N)$ for NN-2 respectively.
The real-valued output is of dimension $4N$ to encode the $M=2$ different complex-valued conditional probability amplitude on $N$ sites.

Similarly, convolutional neural networks (CNNs) are defined by a recursion relation at each layer
\begin{equation}
    {z}^{(l)} = g({W}^{(l)} \ast {z}^{(l-1)} + {b}^{(l)}),
\end{equation}
where now we replace the matrix multiplication with the convolution operation $*$.
To be explicit, we can write it as
\begin{equation}
    {z}^{(l)}_{c,j} = g( \sum_{c'=1}^{\alpha N}\sum_{k=1}^{N_F} {W}^{(l)}_{c,c',k} {z}^{(l-1)}_{c',(j+k-N_F)} + {b}^{(l)}_{c,j}),
\end{equation}
where $N_F$ denotes the filter size.
We consider two different setups of  global connectivity $N_F = N$ and local connectivity $N_F < N$.
We pad the input with zeros to keep the width of the network fixed for simplicity and only change the number of channels at each layer.
For shallow networks, we consider CNNs up to three hidden layers and denote them as CNN-[depth].
The number of channels are $(2,\alpha, 4)$ for CNN-1, $(2,\alpha,\alpha, 4)$ for CNN-2, and $(2,\alpha, \alpha, \alpha, 4)$ for CNN-3.
The output layer is with 4 channels to represent the $M=2$ different complex-valued conditional probability amplitudes.
All shallow NNs and CNNs have ReLU as activation function, i.e. $g(z) = \text{max}(0, z)$.

To turn the networks into networks satisfying autoregressive properties, the connectivity is restricted by masking the weights.
For example, the type-A masking makes sure the $z^{(l)}_i$ does not depend on $z^{(l-1)}_i$ and component afterwards by restricting the sum to $   {z}^{(l)}_i = g( \sum_{j<i} {W}^{(l)}_{i,j}{z}^{(l-1)}_j + {b}^{(l)}_i  ) $.
The type-B masking makes sure the $z^{(l)}_i$ does not depend on $z^{(l-1)}_{i+1}$ and component afterwards by restricting the sum to $   {z}^{(l)}_i = g( \sum_{j\leq i} {W}^{(l)}_{i,j}{z}^{(l-1)}_j + {b}^{(l)}_i  ) $.
Combining the type-A masking at the first layer and type-B masking at all layers afterward, we ensure the autoregressive properties of the network.
For more detail about autoregressive models see Appendix~\ref{appendix: NAQS} and see Fig.~\ref{fig: ansatz} for a graphical illustration.

\begin{figure}[t!]
\centering
\includegraphics[width =\columnwidth]{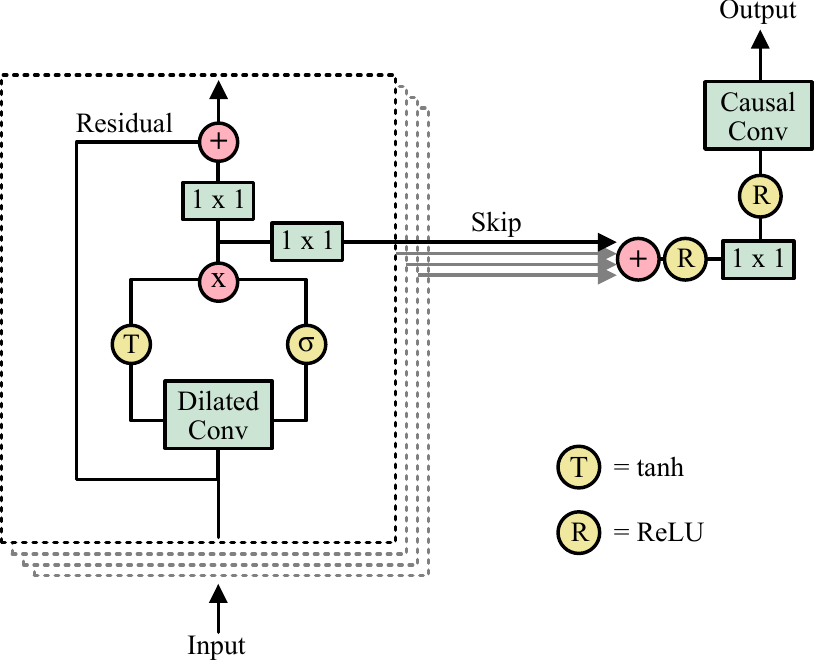}
\caption{
The building block and network architecture of WaveNet-like network adapted from~\cite{oord2016wavenet}.
The building block is indicated by the dotted box.
We repeat the building block to increase the depth.
Except for the first and the last block, the output of the residual connection is the input of the next following block.
The Gated-CNN is a simplification of the WaveNet without skip-connections and the following layers. Additionally, it uses convolution instead of dilated convolution.
}\label{fig: wavenet}
\end{figure}

We consider two different deep neural networks.
In Fig.~\ref{fig: wavenet}, we show the network adapted from the WaveNet~\cite{oord2016wavenet}.
The WaveNet uses dilated convolution, which is the convolution applying over a larger area by skipping input values.
The stacked dilated convolutions with exponential growing dilation factors enable exponential growing receptive field (causal cone).
Similar to the WaveNet, we double the dilation factors at each layer and repeat again once it reach the system size, e.g. $1,2,4,8,16,1,2,4,\cdots$ for $N=20$.

Following the dilated convolution is the gated activation unit~\cite{oord2016conditional},
\begin{equation}
    z = \text{tanh}(W^{f} * x ) \odot \sigma(W^{g} * x)
\end{equation}
where $\odot$ is the element-wise multiplication.
The $\sigma(\cdot)$ is the sigmoid activation function and should not be confused with the local computation basis.
The $W^f$ and $W^g$ are the weight for filter and gate respectively.

After the gated activation unit, there are two special connections.
The residual connections~\cite{he2016deep} are implemented by the $1\times 1$ convolution and the summation with the shortcut from the input of the dilated convolution.
The output of the residual connection is the input of the next block.
The parameterized skip connections are implemented by the $1\times 1$ convolution to form shortcuts from each block directly to the end of repeating blocks, which are then summed together with the final output.
These ``shortcut" connections are the key to training deep neural networks.

In addition to the WaveNet, we consider a simple deep CNN, Gated-CNN, where we remove the skip connections and use only convolution instead of the dilated convolution.
Without the dilated convolution, the receptive field of a Gated-CNN grows linear with depth.
For the receptive field to cover the system size $N=20$, we consider Gated-CNNs with $10$ to $20$ blocks.
As for WaveNets, we consider having $6$ to $14$ blocks.
While WaveNets include one block of $1\times 1$ convolution before the final causal convolution, the output of the Gated-CNN blocks is followed directly by the causal convolution. 
The design choice is arbitrary and turns out not to affect the scaling result.
Similar to shallow CNNs, the final output of both networks has 4 channels to represent the $M=2$ different complex-valued conditional probability amplitudes.

The code implementation based on TensorFlow~\mbox{\cite{tensorflow2015-whitepaper}} is available~\mbox{\cite{GitHub2021}} and the hyperparameters and general setup of the optimization can be found in Appendix~\mbox{\ref{appendix: training_setup}}.

\section{Review of Neural Autoregressive Quantum
States \label{appendix: NAQS}}

A joint probability distribution $P$ over $N$ discrete random variables $x_i$ can be factorized as the multiplication of conditional probabilities.
\begin{align*}
    P(x_1, x_2,\ldots,x_N) &= P(x_1) \times P(x_2 | x_1 ) \times P(x_3 | x_1, x_2) \ldots \\
    &= \prod_{i=1}^{N}p(x_i|x_1,x_2,\ldots,x_{i-1})\\
    &= \prod_{i=1}^{N}p(x_i|{\bf x}_{<i})
\end{align*}
The decomposition imposes an ordering over the random variables and allows for exact sampling.
However, the exact representation of conditional probabilities also scales exponentially with the number of random variables.
Autoregressive models are parameterized functions approximating the conditional probabilities.

One important class are neural autoregressive models~\cite{bengio2000modeling,uria2016neural}.
By utilizing neural networks to approximate the conditional probabilities, it allows exact and efficient sampling and inference.
It has been applied to studying statistical physics in the pioneering work~\cite{wu2019solving} and in~\cite{liu2021solving}.

In principle, one can approximate each conditional probability with a different neural network,
\begin{equation*}
P_\text{AR}= f(x_1) \times g(x_2|x_1) \times h(x_3 | x_1, x_2) \times \ldots
\end{equation*}
A simpler and common approach is to approximate all $N$ conditional probability by one neural network with $N$ inputs and $N$ outputs,
\begin{equation*}
P_\text{AR}= f^{[1]}(x_1) \times f^{[2]}(x_2|x_1) \times f^{[3]}(x_3 | x_1, x_2) \times \ldots
\end{equation*}
The network can no longer be fully-connected, but with connectivity satisfying the autoregressive property.
That is $f^{[2]}$ can only depend on input $x_1$ and $f^{[3]}$ only on inputs $x_1,x_2$ and so on.
It is common to enforce this by zero-masking the original weights $w$.
Equivalently, we say we have masked weight multiplication or masked convolution (see Fig.~\ref{fig: ansatz}).

Following the same principle, a probability amplitude can be rewritten as a product of conditional probability amplitudes~\cite{albareda2014correlated}.
\begin{equation}
\Psi_{\sigma_1 \sigma_2 \cdots \sigma_N} = \Psi_{\sigma_1} \Psi_{\sigma_2|\sigma_1} \cdots \Psi_{\sigma_N|\sigma_1 \cdots \sigma_{N-1}}
\end{equation}
Neural autoregressive quantum states (NAQS)~\cite{sharir2020deep} approximate the conditional probability amplitudes by neural networks.
\begin{equation}
    \Psi^{NAQS}_{\sigma_1 \sigma_2 \cdots \sigma_N} 
    =  f_\text{NN}^{[1]}(\sigma_1) f_\text{NN}^{[2]}(\sigma_2|\sigma_1) \cdots f_\text{NN}^{[N]}(\sigma_N|\sigma_1 \cdots \sigma_{N-1}) 
\end{equation}
In the following, we denote $f^{[i]}_\text{NN}$ as $\psi_i$.

It is shown that if each conditional probability amplitude is normalized $\sum_{\sigma_i} |\Psi_{\sigma_i| \sigma_{j<i}} |^2 = 1$, then the full wavefunction is normalized $|\Psi_{\sigma_1 \sigma_2 \cdots \sigma_N}|^2=1$.
Therefore, we can normalize the full wavefunction by normalizing each conditional probability amplitude.
Suppose the local basis is of dimension $M$, each of the $N$ outputs is actually a complex vector $\bm{v_i}=(v_{i,1},v_{i,2},\ldots, v_{i,M})$, which represents the logarithm of the conditional probability amplitude at site $i$.
\begin{equation}
\psi_i(\sigma_i | \sigma_1,\ldots,\sigma_{i-1}) = \frac{e^{v_{i,\sigma_i}}}{ \sum_{\sigma_i'} | e^{v_{i,\sigma_i'}} |^2 }
\end{equation}
gives the normalized conditional probability amplitudes.
Furthermore, each $|\psi_i|^2$ gives a valid conditional probability induced by Born's rule.
This means a NAQS contains a neural autoregressive model, which permits efficient and exact sampling.

\section{Cost Functions and Stochastic Gradient Descent \label{appendix: cost}}

In this section, we give a brief review of the setup of supervised learning problems and discuss the difference between it and the general optimization problem.
We further give an overview of different cost functions for variational states approximation and the resulting properties. The discussion is general and applies to general variational wavefunctions.

Supervised learning is a minimization problem to learn a function mapping from $X$ to $Y$.
Given the data $\{ x, y\}^N$, where the data pair $\{x, y\} \sim P(x, y)$ are sampled from an unknown probability distribution $P(x, y)$, and the parameterized function $f:x\in X\rightarrow f(x) \in Y$ with parameters $w$, we would like to minimize the cost function $\mathcal{R}(f)$ with respect to parameters $w$. $\mathcal{R}(f)$ is defined as
\begin{equation} \label{risk_def}
    \mathcal{R}(f) = \int \mathcal{L}(f(x;w), y) dP(x, y)
\end{equation}
where $\mathcal{L}(f(x;w), y)$ is the loss function measuring the difference between $f(x;w)$ and $y$.
In general, the expression cannot be evaluated.
Instead, the supervised learning framework, a.k.a. empirical risk minimization, relies on minimizing the cost from the sampling (empirical risk) $\hat{\mathcal{R}}(f)$,
\begin{equation}
    \hat{\mathcal{R}}(f) = \sum_{\{x,y\}\sim P(x,y)} \mathcal{L}(f(x;w), y) .
\end{equation}
One can compute the full batch gradient based on the empirical risk and perform a gradient descent algorithm.
In modern applications, often even the empirical risk itself is too expensive to evaluate at all steps given the large size of the dataset.
One turns to compute the mini-batch stochastic gradient $\widehat{\nabla \mathcal{R}}= \sum_i \nabla_w \mathcal{L}(f(x_i;w), y_i)$ where the sum is taken over the mini batch samples.
In terms of deep learning, the supervised learning framework fits nicely with the stochastic gradient descent algorithm.

Another type of optimization problem is to minimize the cost function $\mathcal{R}$, which is not in the form as in Eq.~\eqref{risk_def}, i.e. an integral or a sum of the loss function $\mathcal{L}$.
The cost function $\mathcal{R}$ itself is the thing we want to optimize.
In this case, gradient-based algorithms would require computing the gradient $\nabla_w \mathcal{R}$.
In certain cases, the gradient $\nabla_w \mathcal{R}$ itself cannot be computed exactly and approximation is required.
Moreover, the gradient $\nabla_w \mathcal{R}$ might contain intractable terms as integral or summation, which can be approximated by sampling.
In this case, the sampled ``stochastic gradient" has quite different properties comparing to the previous one in the setup of supervised learning.
More precisely, \emph{the stochastic gradient is unbiased when the expectation value over the sampling of this stochastic gradient equal to the true gradient}, i.e.
\begin{equation*}
    \nabla \mathcal{R} = \mathbb{E}[ \widehat{\nabla \mathcal{R}} ]
\end{equation*}
The stochastic gradient is \emph{biased} if the expectation value of the sampled stochastic gradient does not equal to the true gradient.
Stochastic gradients in supervised learning setup are unbiased while in the general optimization problem they are not.
With this in mind, we can distinguish the cost function for variational state approximation by whether it has the form of Eq.~\eqref{risk_def}.

In terms of the variational state approximation, given the target wavefunction $| \Phi \rangle$, the variational wavefunction $| \Psi\rangle$ and the dimension of the many-body Hilbert space $D_\mathcal{H}$. here we review different possible cost functions that work for both unnormalized wavefunctions and normalized wavefunctions introduced in previous works.

In \cite{cai2018approximating,kochkov2018variational}, the $\mathcal{L}_2$ distance of the probability amplitudes $\mathcal{R}_\text{MSE}$ are considered,
\begin{align*}
\mathcal{R}_\text{MSE} &= \frac{1}{D_\mathcal{H}} \sum_{i}^{D_\mathcal{H}} |\Psi(X_i) - \Phi(X_i) |^2 \\
&= \frac{1}{D_\mathcal{H}} \sum_{i}^{D_\mathcal{H}} P_{\Phi}(X_i) \frac{ |\Psi(X_i) - \Phi(X_i) |^2 }{ \lvert \Phi(X_i) \rvert^2 }.
\end{align*}
Similarly, in Ref.~\onlinecite{westerhout2020generalization}, the weighted $\mathcal{L}_2$ distance of log of the probability amplitudes is considered,
\begin{align*}     
\mathcal{R}_\text{weighted MSE\ log} &= \frac{1}{D_\mathcal{H}} \sum_{i}^{D_\mathcal{H}} P_{\Phi}(X_i) \Big | \log\Psi(X_i) - \log\Phi(X_i) \Big |^2
\end{align*}
while in \cite{cai2018approximating} a uniform sampling scheme is considered, and in Ref.~\onlinecite{kochkov2018variational,westerhout2020generalization} the samples are taken with respect to $P_\Phi$. These cost functions are unbiased but with high variance. Furthermore, the problem is posed as a regression problem instead of taking into account the probability nature of the problem.

A different setup is considered in Ref.~\onlinecite{jonsson2018neural} with the negative log fidelity as the cost function.
\begin{align*}
        &\mathcal{R}_\text{neg. log fidelity} =-\log\Big[ \frac{\braket{\Psi|\Phi}\braket{\Phi|\Psi}}{\braket{\Psi|\Psi}\braket{\Phi|\Phi}} \Big] \\
        &=- \log\Bigg[ \frac{\Big( \sum_{i}^{D_\mathcal{H}}  \Psi^*(X_i)\Phi(X_i) \Big) \Big( \sum_{i}^{D_\mathcal{H}} \Phi^*(X_i)\Psi(X_i) \Big) }{
        \Big( \sum_{i}^{D_\mathcal{H}}  \Psi^*(X_i)\Psi(X_i) \Big) \Big( \sum_{i}^{D_\mathcal{H}} \Phi^*(X_i)\Phi(X_i)\Big)} \Bigg]
\end{align*}
This cost function and the corresponding gradient are in the form that could be estimated by stochastic sampling. However, the resulting stochastic gradient is biased. In practice, this cost function works well as long as the batch size is taken large enough.
In this work, we use directly the negative fidelity as the cost function when optimizing the RBMQS.
\begin{align*}
    \mathcal{R}_\text{neg. fidelity} &= -\frac{(\sum_i \Psi_i^* \Phi_i)(\sum_i \Psi_i \Phi^*_i) }{(\sum_i \Psi_i^* \Psi_i)(\sum_i \Phi_i^* \Phi_i)} \\
    &= -\frac{ \mathbb{E}_{i\sim P_\Phi}[\Psi^*_i/\Phi_i^*] \mathbb{E}_{i\sim P_\Phi}[\Psi_i/\Phi_i] }{\mathbb{E}_{i\sim P_\Phi}[ |\Psi_i|^2 / |\Phi_i|^2 ] }
\end{align*}
With the stochastic estimate of the fidelity, the biased stochastic gradient is then obtained from the gradient evaluated by automatic differentiation framework.

In this work, we consider two cost functions that work only for normalized wavefunction, which give unbiased stochastic gradient with low variance. In Section~\ref{subsec: supervised_learning}, we mainly focus on the cost function
\begin{equation}
    \mathcal{R}_\text{joint} = \mathcal{R}_\text{KL} + \mathcal{R}_{\theta}.
\end{equation}
An alternative cost function is the real part of the negative overlap,
\begin{align*}
    \mathcal{R}_\text{neg. overlap} 
    &= -\sum_i^{D_\mathcal{H}} \text{Re} \left [  \braket{\Phi|X_i} \braket{X_i|\Psi } \right ]\\
    &= -\sum_i P_\Phi(X_i) \text{Re} \left [  \frac{\braket{X_i|\Psi }}{\braket{X_i|\Phi }} \right]
\end{align*}
Minimizing the real part $\text{Re} \left [  \braket{\Phi|\Psi} \right ]$ is equivalent to minimizing $\lvert \braket{\Phi|\Psi} \rvert$ because we can always absorb a phase factor into our complex-valued wavefunction.
Minimizing only the real part gives the form of expectation value and again we can apply the stochastic gradient descent method.
This, however, cannot apply to unnormalized wavefunction because of the intractable normalization constant.

The fidelity $\mathcal{F}$ is the square of the overlap, $\mathcal{F} = \lvert \braket{\Phi|\Psi} \rvert^2$. It may be tempting to maximize the fidelity $\mathcal{F}$ instead by the sampling estimate of the overlap. However, again we encounter the problem of a biased gradient.
Although, the fidelity estimate cannot serve as a cost function. We can still evaluate it as a check of the validity of the final result.

Finally, we show that these two cost functions are equivalent to the first order.
\begin{align*}    
\mathcal{R}_\text{neg. overlap} &= -\sum_i P_\Phi(X_i) \text{Re} \left [  \frac{\braket{X_i|\Psi }}{\braket{X_i|\Phi }} \right]\\
&= -\sum_i P_\Phi(X_i) \text{Re} \left [  \frac{e^{r^\Psi_i+ i\theta^\Psi_i}}{e^{r^\Phi_i+ i\theta^\Phi_i}} \right]\\
&= -\sum_i P_\Phi(X_i) \text{Re} \left [  e^{(r^\Psi_i-r^\Phi_i)+ i(\theta^\Psi_i-\theta^\Phi_i)} \right] \\
&\approx -\sum_i P_\Phi(X_i)  \left [  (1+(r^\Psi_i-r^\Phi_i)) \right(1+ \frac{(\Delta \theta_i)^2}{2} \left) \right] \\
&\approx \sum_i P_\Phi(X_i)  \left [-1 + (r^\Phi_i-r^\Psi_i) - \frac{(\Delta \theta_i)^2}{2} \right] \\
&\approx -1 + \frac{1}{2}(\mathcal{R}_\text{KL} + \mathcal{R}_{\theta})  \\
\end{align*}
In practice, we observe slightly different dynamics at the beginning of the learning process.

\section{Setup of Optimizations \label{appendix: training_setup}}

Below, we first discuss the common setup of optimizations for both NAQS and MPS.
\begin{itemize}
    \item 
    \emph{Optimizer and batch size:} We consider Adam optimizer and a batch size of $512$.

    \item 
    \emph{Convergence criterion:}
    The convergence criterion is based on monitoring the error in fidelity.
    We set the criterion to be either if the error were below $10^{-4}$ or if average error of the latest $500$ batches were higher than the average error of the latest $5000$ to $2500$ batches.
    In addition, there is a break condition set by the maximal number of steps.

    \item
    \emph{Maximal number of steps:} 
    We set the maximal number of steps to be $3\times 10^{5}$, which is almost never reached.
    The typical number of steps are below $10^5$.

    \item
    \emph{Number of runs}:
    For each data point, we run the optimization with random initialization $4$ times and report the best result.
    
\end{itemize}

\emph{Matrix Product States:} The MPS considered are complex-valued.
We start out with a fix learning rate $10^{-2}$ until convergence. Then, we repeat the procedure with the learning rate decreased to $10^{-3}$ and $10^{-4}$.

Notice that the canonical (isometric) form or normalization of MPS is not enforced during the optimization.
In fact, we observed that the result is better than the result enforcing the isometric form using Riemannian optimization~\mbox{\cite{hauru2021riemannian,luchnikov2020qgopt}}.

\emph{Neural Network:}
All the networks considered are parameterized by real-valued weights and biases.
We start out with a fix  learning rate $10^{-3}$ until the convergence criterion is met. Then, we decrease the learning rate to $10^{-4}$ and continue again the optimization until the convergence.

Here, we provide the setup of using NetKet to perform state approximation with RBMQS.
 
\emph{NetKet:}
\begin{itemize}
    \item
    \emph{Version}:
    2.1
    
    \item 
    \emph{Optimizer and batch size}:
    stochastic reconfiguration and stochastic gradient descent. The batch size is $1000$. We take the best result from different step sizes: $1\times 10^{-1}, 3\times 10^{-2},  1\times 10^{-2}$.
    
    \item 
    \emph{Convergence criterion:}
    The convergence criterion is based on monitoring the average error in fidelity.
    We set the criterion to be either if the average error were below $10^{-4}$ or if average error of the latest $50$ batches were close to the average error of the latest $100$ to $50$ batches with a relative error $3\times 10^{-6}$ and absolute error $3\times 10^{-7}$.
    In addition, there is a break condition set by the maximal number of steps.

    \item
    \emph{Maximal number of steps:}
    We start checking for convergence after 5000 iterations, and set a maximum iteration of $10000$ which is almost never reached.

    \item
    \emph{Number of runs}:
    For each data point, we run the optimization with random initialization $4$ times and report the best result.
\end{itemize}

\section{Additional Data \label{appendix: data}}

\begin{figure}[t]
\centering
\includegraphics[width =\columnwidth]{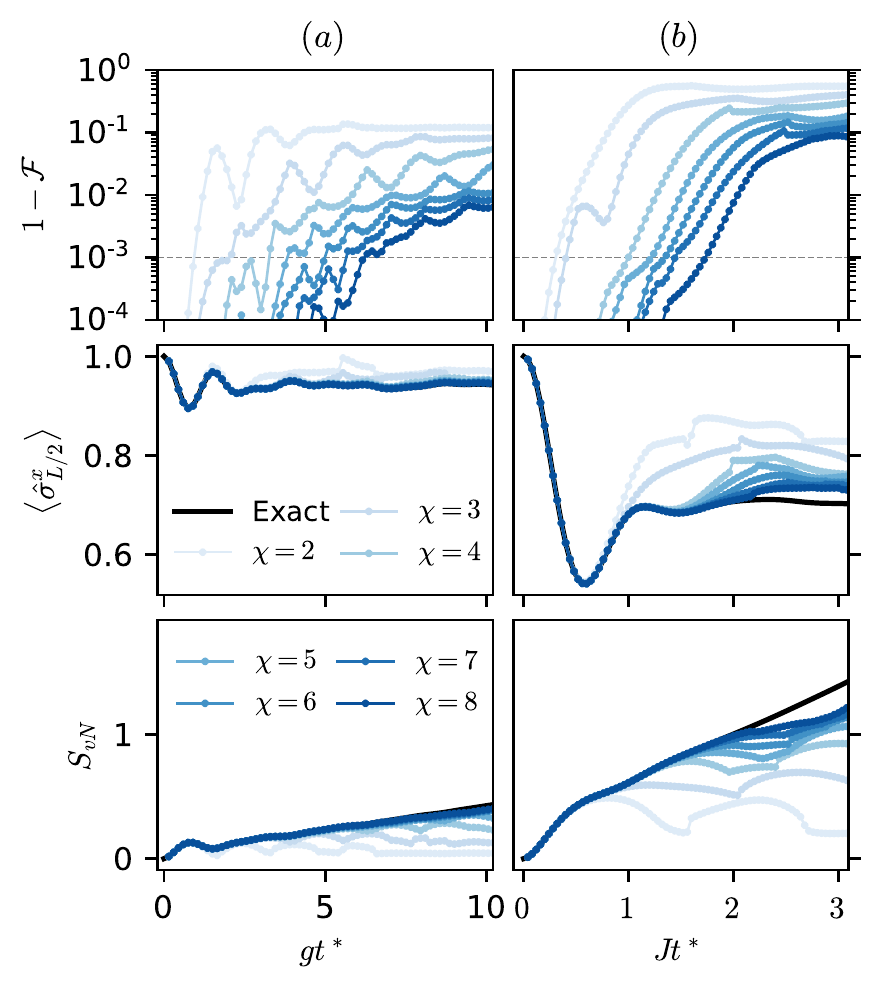}
\caption{MPS with bond dimension $\chi$ approximating the exact time-evolved states following a quantum quench from an initial paramagnetic product state.
The MPS are obtained by iteratively minimizing the $\mathcal{L}_2$ distance.
The quenched Hamiltonian is given in Eq.~\eqref{eq: Hamiltonian} for a chain of length $N=20$ with (a) quantum Ising model in the paramagnetic phase with weak longitudinal field ($g=3$, $h=0.1$, $k=0$) and (b) close to critical point with interacting term ($g=1$, $h=0$, $k=0.25$).
\label{fig: MPS_ALS}}
\end{figure}

In Fig.~\mbox{\ref{fig: MPS_ALS}}, we include the result of MPS approximating the target state by the alternating least square algorithm, which iteratively minimizes the $\mathcal{L}_2$ distance of the quantum states~\mbox{\cite{schollwock2011density}} and yield slightly better result than SGD optimization with supervised learning as shown in Fig.~\ref{fig: MPS_data}.

\begin{figure}[h]
\centering
\includegraphics[width =\columnwidth]{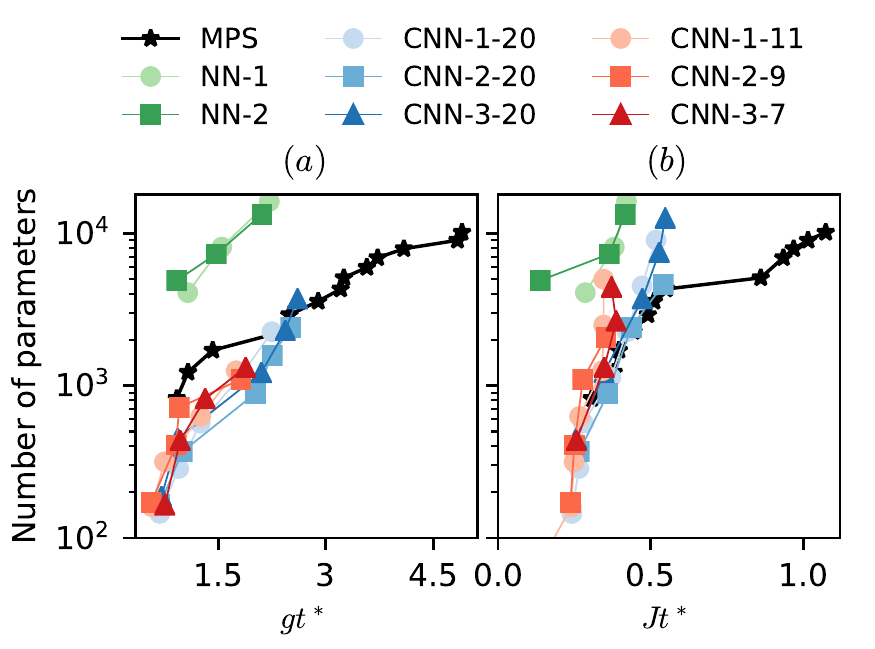}
\caption{
The number of parameters required to reach time $t^*$ for following a quantum quench for different approximation schemes.
The quenched Hamiltonian is similar to Eq.~\eqref{eq: Hamiltonian} but with PBC on a chain of length $N=20$ with (a) quantum Ising model in paramagnetic phase with weak longitudinal field ($g=3$, $h=0.1$, $k=0$) and (b) close to critical point with interacting term ($g=1$, $h=0$, $k=0.25$).
The black line shows the results for MPS.
For NAQS, we denote each combination as (network type)-(number of hidden layers)-(filter size).
The circles, squares, triangles represents 1,2,3-layer networks respectively.
The gradient colors describe same networks of different number of layers.
NNs (Blue); CNNs with global connectivity (Green-Blue); CNNs with local connectivity (Red).
}\label{fig: para_time_scaling_PBC}
\end{figure}

In Fig.~\mbox{\ref{fig: para_time_scaling_PBC}}, we include the results with shallow NAQS for periodic boundary condition (PBC) and plot the number of parameters of the network against the obtained accessible time $t^*$.

\begin{figure}[h]
\centering
\includegraphics[width=\columnwidth]{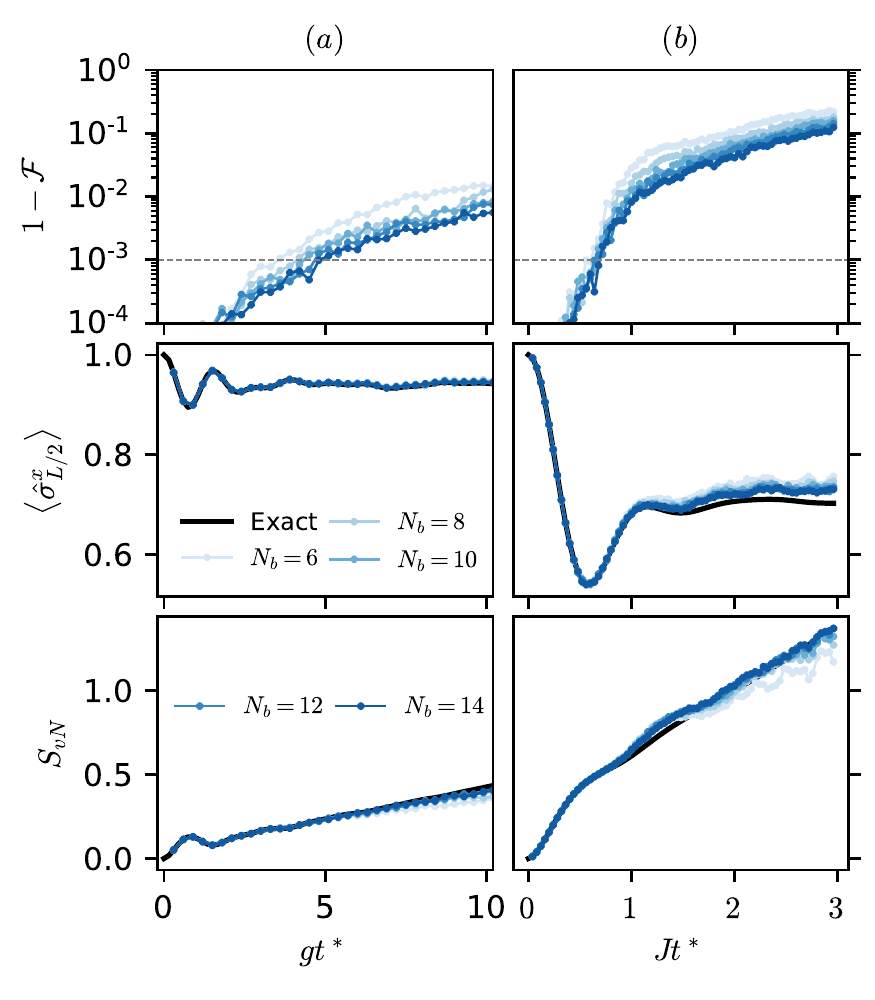}
\caption{WaveNet NAQS of different depths, i.e. number of blocks $N_b$, approximating the exact time-evolved states following a quantum quench from an initial paramagnetic product state.
The network is of width $\alpha=8$.
The quenched Hamiltonian is given in Eq.~\eqref{eq: Hamiltonian} for a chain of length $N=20$ with (a) quantum Ising model in the paramagnetic phase with weak longitudinal field ($g=3$, $h=0.1$, $k=0$) and (b) close to critical point with interacting term ($g=1$, $h=0$, $k=0.25$).
\label{fig: deep_NN_data_wavenet}}
\end{figure}

In Fig.~\ref{fig: deep_NN_data_wavenet}, we show the results for WaveNets of width $\alpha=8$.

\begin{figure}[h]
\centering
\includegraphics[width=\columnwidth]{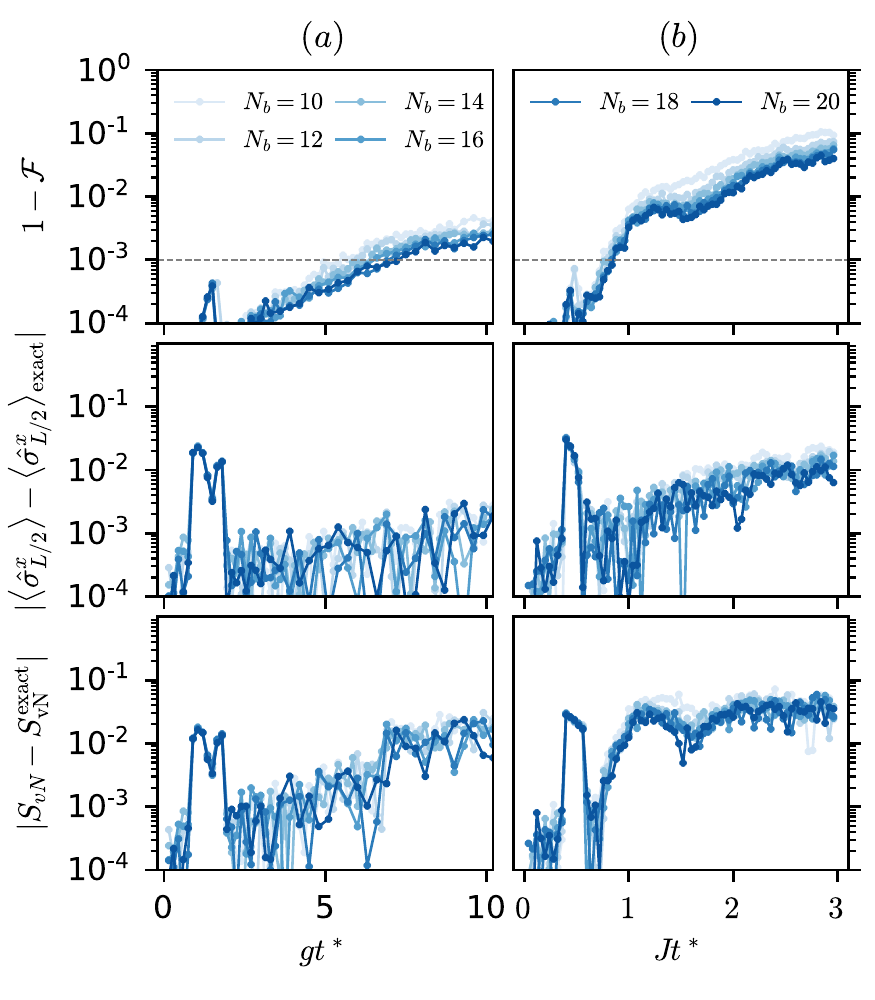}
\caption{The errors of Gated-CNNs NAQS of different depth, i.e. number of blocks $N_b$, approximating the exact time-evolved states following a quantum quench from an initial paramagnetic product state.
The network is of width $\alpha=12$.
The quenched Hamiltonian is given in Eq.~\eqref{eq: Hamiltonian} for a chain of length $N=20$ with (a) quantum Ising model in the paramagnetic phase with weak longitudinal field ($g=3$, $h=0.1$, $k=0$) and (b) close to critical point with interacting term ($g=1$, $h=0$, $k=0.25$).
\label{fig: deep_NN_data_gatedCNN_error}}
\end{figure}

In Fig.~\mbox{\ref{fig: deep_NN_data_gatedCNN_error}}, we plot the errors of the results for Gated-CNNs of width $\alpha=12$ and different depths.
We see the errors in $\langle \hat{\sigma}^x_{L/2}\rangle$ expectation value and half-chain entanglement entropy $S_\text{vN}$ follows similar tendency as of the error in fidelity.
Interestingly, we observe a peak in both errors at a relatively short time before the network reaches its accessible time $t^*$.

\begin{figure}[h]
\centering
\includegraphics[width =\columnwidth]{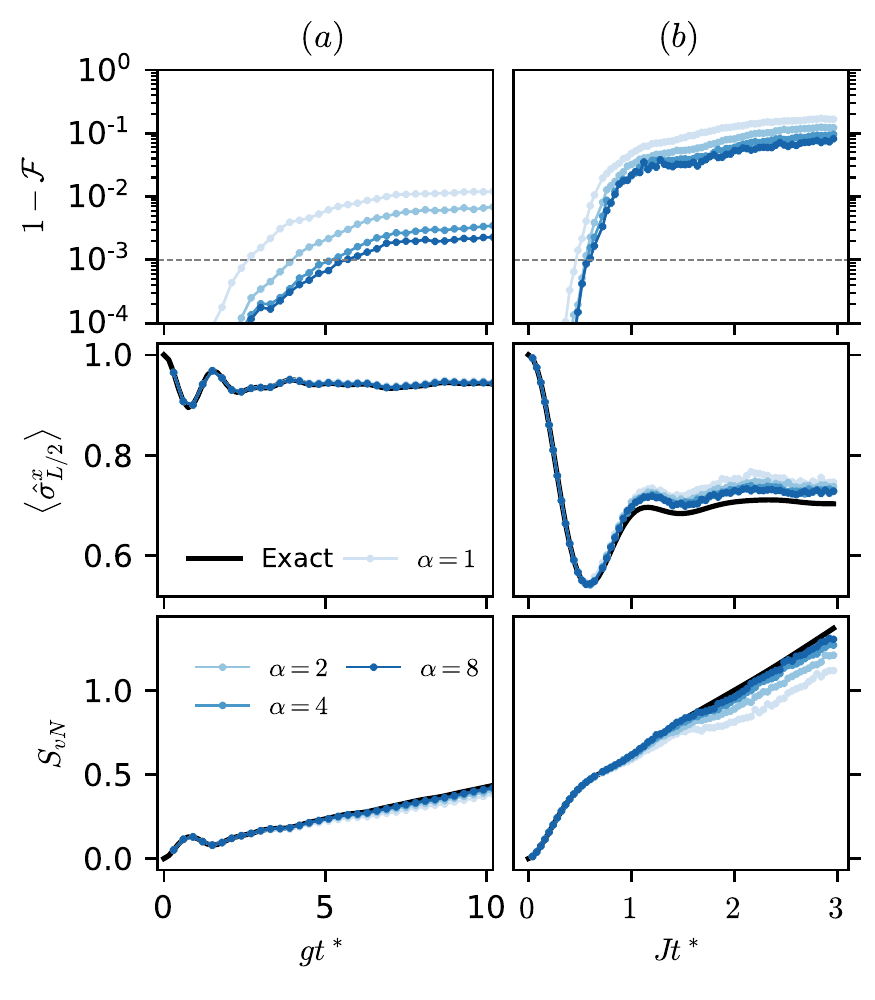}
\caption{ RBMQS of different widths, i.e. $\alpha=N_\text{hidden}/N$, approximating the exact time-evolved states following a quantum quench from an initial paramagnetic product state.
The quenched Hamiltonian is given in Eq.~\eqref{eq: Hamiltonian} for a chain of length $N=20$ with (a) quantum Ising model in the paramagnetic phase with weak longitudinal field ($g=3$, $h=0.1$, $k=0$) and (b) close to critical point with interacting term ($g=1$, $h=0$, $k=0.25$).
\label{fig: RBM}}
\end{figure}

In Fig.~\mbox{\ref{fig: RBM}}, we show the results for RBMQS.

\begin{figure*}[h]
\centering
\includegraphics[width=0.8\textwidth]{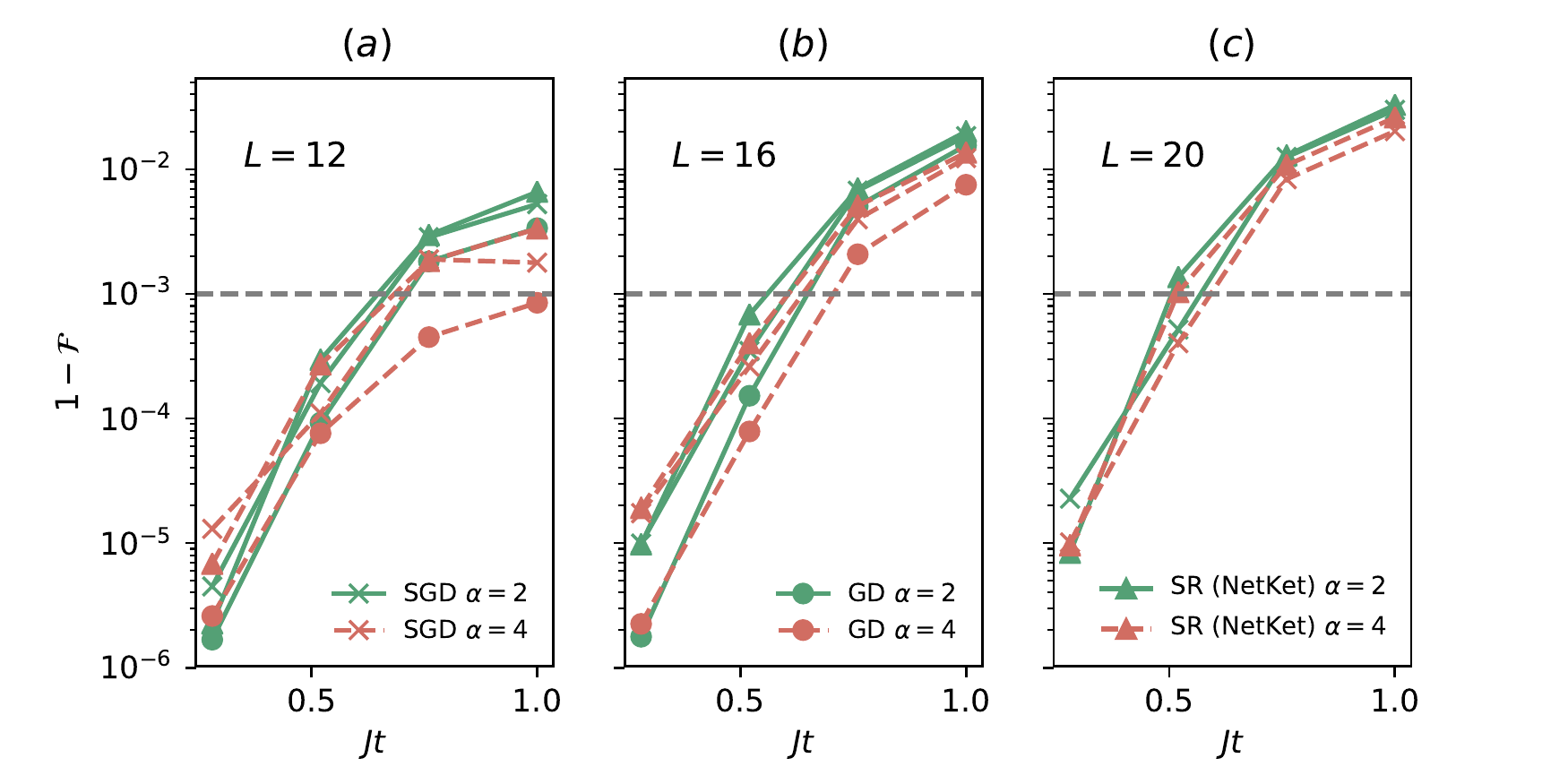}
\caption{ Comparison of results obtained from different optimizers and implementations for RBMQS of two different widths $\alpha=2,4$ approximating the exact time-evolved states following a quantum quench from an initial paramagnetic product state.
The quenched Hamiltonian is given in Eq.~\eqref{eq: Hamiltonian} for a chain of length $N=12,16,20$ with quantum Ising model close to critical point with interacting term ($g=1$, $h=0$, $k=0.25$).
\label{fig: RBMQS_consistency}}
\end{figure*}

In Fig,~\mbox{\ref{fig: RBMQS_consistency}}, we compare the results for state approximation by RBMQS using SGD and exact gradient optimization and the SGD with stochastic reconfiguration from the NetKet~\mbox{\cite{carleo2019netket}} implementation.

\begin{figure*}[h]
\centering
\includegraphics[width=0.8\textwidth]{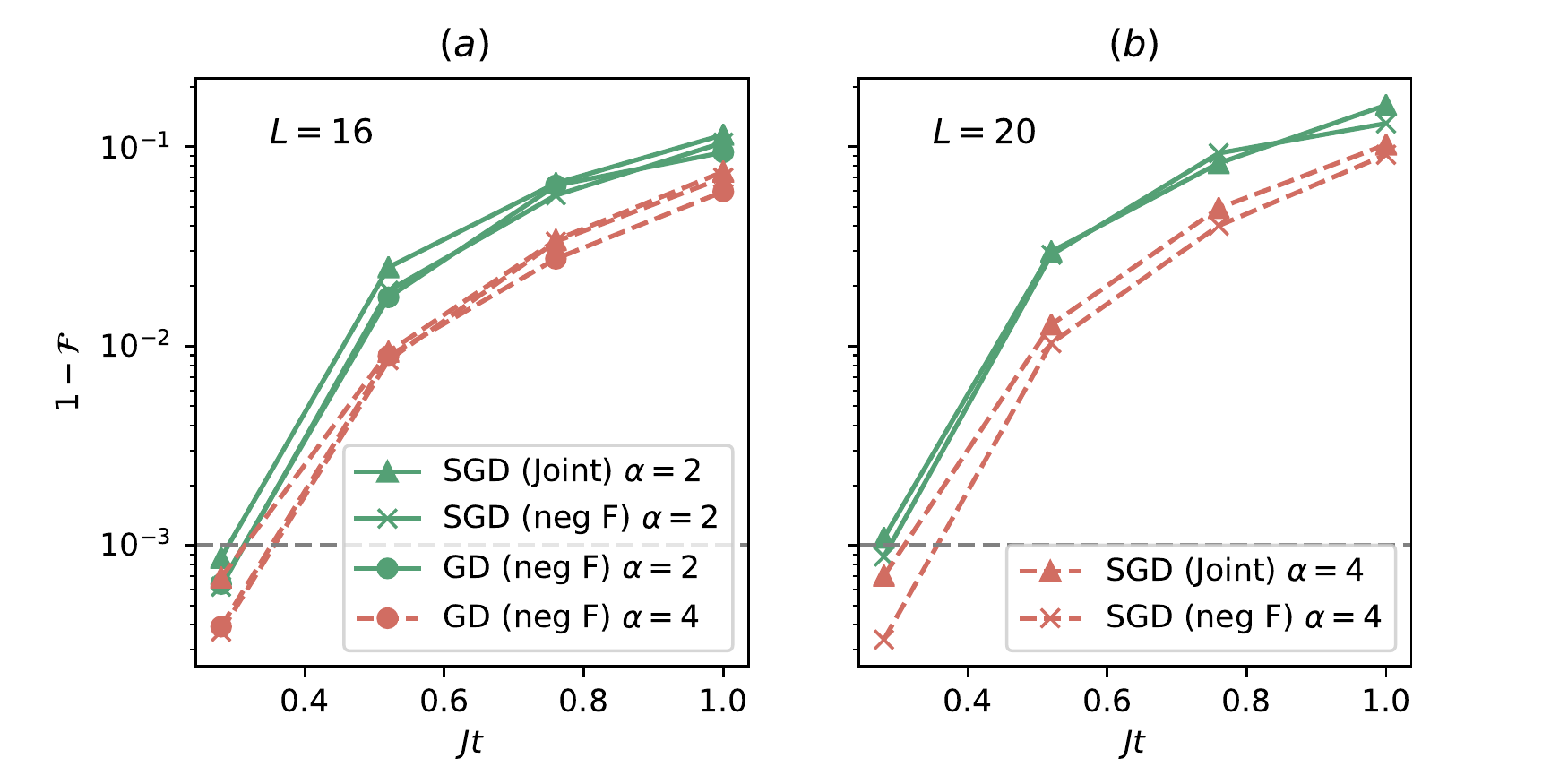}
\caption{ Comparison of results obtained by different optimizers and cost functions for one-hidden layer NAQS of two different widths $\alpha=2,4$ approximating the exact time-evolved states following a quantum quench from an initial paramagnetic product state.
The quenched Hamiltonian is given in Eq.~\eqref{eq: Hamiltonian} for a chain of length $N=12,16,20$ with quantum Ising model close to critical point with interacting term ($g=1$, $h=0$, $k=0.25$).
\label{fig: MADE_consistency}}
\end{figure*}

In Fig,~\mbox{\ref{fig: MADE_consistency}}, we compare the results for state approximation by one-hidden layer NAQS using SGD and GD over negative fidelity and using SGD over joint cost function.

\clearpage

\bibliography{apssamp}%

\end{document}